\documentclass[11pt,a4paper]{article}
    \usepackage{jheppub}
\newcommand{\Array}[2]{\left(\begin{array}{#1}#2\end{array}\right)}

\title{The Minimal Flavor Structure of Quarks and Leptons}
\author[a]{Ying Zhang}
\affiliation[a]{Institute of Theoretical Physics, School of Physics, 
	\\
	Xi'an Jiaotong University, Xi'an, 710049, China}
\emailAdd{hepzhy@mail.xjtu.edu.cn}

\abstract{A flavor structure with minimal parameters is proposed to address the fermion mass hierarchy and flavor mixing for  quarks and leptons. Yukawa interaction is reconstructed 
 in a new basis to show a homological flat structure for up-type quarks, down-type quarks, charged leptons and  Dirac neutrinos. A $SO(2)_{LR}^f$ flavor symmetry is found from the hierarchy masses of quarks and leptons, which dominated CKM mixing for quarks and PMNS for leptons. 
Since the minimal flavor structure  successfully addresses CKM and PMNS even in the mass hierarchy limit, mass hierarchy and flavor mixing are two independent questions. As a prediction, a sum rule on the mixing angles and CP violation phase is  suggested, which explains the smallness of $s_{13}$ as a natural result of  the mass hierarchy.  
Generalizing the flat structure to quarks and leptons, a unified Yukawa interaction is achieved for all  fermions with only a single coupling.
}

\begin{document}
\maketitle

\tableofcontents

\newpage
\section{Overview}\label{sec.Movt}

The Standard Mode (SM) has successfully and clearly described $SU(3)_c\times SU(2)_L\times U(1)_Y$ gauge interactions.
Strong, weak and hypercharge interactions are controlled by three simple couplings.
The fermions in the SM with the same quantum number are divided into three families (generations), which are only different in their masses. Through gauge symmetry, each family has non-distinguished gauge interactions.

However, the SM has not given a satisfactory structure in terms of flavor. In the SM, fermion flavor breaking is induced in the Yukawa term. Flavor-dependent Yukawa couplings produce a fermion mass matrix after the EW symmetry is broken. Due to different diagonalizing transformations of the up-type and down-type quark mass matrices, a manifestation of flavor breaking in the quark sector is transmitted to the weakly charged current interaction, which is known as CKM mixing. Similarly, lepton flavor breaking also involves the PMNS mixing matrix in charged current weak interactions. Thus, mass difference and nontrivial flavor mixing become two aspects of flavor manifestations of quarks and leptons. They both occur from flavor-dependent Yukawa couplings. 
However, the SM cannot provide detailed information on these coupling values or structures. Ambiguous Yukawa interactions in the SM involve flavor physics puzzles, such as
	(1) Where does the fermion hierarchical mass comes from?
	(2) Is there a common structure regarding the CKM mixing of quarks and PMNS mixing of leptons?
	(3) Are hierarchical mass and flavor mixing two independent questions?
To answer these questions, many interesting mechanisms and models have been proposed; moreover, these mechanisms and models are inspired by the following characteristics of flavors:
\begin{itemize}
\item[(1)] Hierarchal masses. Regarding up-type and down-type quarks and charged leptons, their masses that are arranged as generations have a hierarchal structure: $m_1^f\ll m_2^f\ll m_3^f$ for $f=u,d,e$. Normal-order neutrinos also meet $m_1^\nu \ll m_2^\nu\ll m_3^\nu$.
There are three main kinds of modes that address hierarchy masses:
	\begin{itemize}
		\item $\Array{cc}{0 & \epsilon \\ \epsilon & 1}$ mode. It is well known as the seesaw mechanism \cite{Yanagida1979, Yanagida1979PRD, Mahapatra1980, GavelaJHEP2009}. The lighter mass is generated from the nondiagonal correction $\epsilon$.
		\item $\Array{cc}{\epsilon & 0 \\ 0 & 1}$ mode. Only the heavier family is massive at the tree level. The lighter family mass is generated from some quantum corrections \cite{BalakrishnaPRL1988, BalakrishnaPLB1988, MohantaPRD2022}. The recent work in \cite{2020Weinberg} has tried to generate the second and first families by one-loop and two-loop radiative corrections.
		\item $\Array{cc}{1 & 1 \\ 1 & 1}$ mode. Due to the equal treatment of each flavors, the flat matrix is also called the democratic matrix \cite{HarariPLB1978, KoidePRL1981, KoidePRR1983,XingPRD2000, FritzschPLB2004}. This matrix has eigenvalues $(0,2)$. When some flavor breaking patterns are considered, a small mass for the lighter family can be generated.
	\end{itemize}
These modes can also be generated for three-family hierarchal fermions independently or cooperatively. For example, in \cite{FritzschNPB1979}, a sequential see-saw mechanism with the pattern $$\Array{ccc}{0 & \epsilon_1 & 0 \\ \epsilon_1 & 0 & \epsilon_2\\ 0 & \epsilon_2 & m_{t/b}}$$ was suggested to address quark masses.
		\item[(2)]  Flavor mixing of quarks and leptons. In the experiment, the CKM and PMNS mixing parameters were measured. In contrast to CKM with three small mixing angles, lepton PMNS exhibits two large mixing angles. Some approximate mixing textures, such as the BM and TBM, are suggested for PMNS. The pattern of the mixing matrix has inspired flavor symmetry research (especially in the lepton sector), including discrete symmetries \cite{FeruglioNPB2008, HolthausenPRD2013, ZhouJHEP2020, HabaNPB2006, AltarelliNPB2005} and gauged flavor symmetries \cite{FeruglioRMP2021,GengCPC2018}. Recently, the modular invariances in \cite{LiuJHEP2022} and the symmetry-assisted anarchy method \cite{SFGePLB2018} have also been studied in neutrino mixing patterns. More details on discrete flavor symmetry can be reviewed in \cite{King2016RPP,ZZX2020PR,FeruglioEPJC2015}.
		\item[(3)] Nonvanishing CP violation. In CKM, the CP violation phase has been measured from the $B$ and $K$ decay processes. In PMNS, a value of approximately $3\pi/2$ is commonly suggested for lepton CPV. 
These invoke an assumption on the common origin of quark and lepton CP violations \cite{BrancoRMP2012, BrancoPLB2007, BrancoarXiv}. In \cite{AlvesEPJC2021}, this motivation is addressed from a vacuum phase in a class 2HDM.
	\end{itemize}
	
Nevertheless, new flavor models face multiple challenges. For example, determining the organization principle of flavors from dada and symmetries is difficult. Moreover, the characteristics of the parameters in new models need to be explained. New physics beyond the SM provides additional effects that explain current mass values and/or flavor mixing angles. Regarding additional effects, it is also challenging to find the predicted particles and interactions in future experiments.
These models and mechanisms provide some significant attempts to uncover nature of flavors. Furthermore, there are still some challenges and unknown questions that need to be answered.
Theoretically, hierarchal masses and flavor mixing are two sides of the same question in flavor physics, and they all arise from quark/lepton mass matrices. If treating the mass hierarchy as a good approximation, flavor mixing must provide CKM/PMNS patterns. This means that flavor mixing should come from an independent mechanism on the mass hierarchy.

In addition, quark masses and CKM/PMNS have been successfully measured in experiments. Furthermore, these results have become useful for determining new flavor models. 
When combining the mass spectrum and flavor mixing as a benchmark, many early models have been ruled out by current experimental data.

Therefore, new research must be performed to determine the fermion flavor structure.
The desired flavor structure should have the following properties:
	\begin{itemize}
		\item[(1)] No redundant parameters. Fewer free parameters represent more information that we know. In the quark sector, all flavor parameters include $3+3$ quark masses and 4 CKM parameters. In the lepton sector, the number is the same. We hope to build a description of flavor physics with only $10+10$ parameters for quarks and
leptons. This classification is called the minimal flavor structure (MFS).
		\item[(2)] A general structure independent of the concrete values of quark/lepton masses and mixing parameters. With measurements becoming increasingly precise , the allowed range of experimental values will decrease, and even the center of the value will be shifted. Regarding flavor models, 
an increasing number of corrections must be considered to fill the gap between model predictions and experimental measurements.
Moreover, this structure will challenge the models that are currently used.
Therefore, a desired flavor structure should provide a general mechanism that generates fermion masses and mixing, and this mechanism can remain effectively independent of concrete experimental values. It requires that MFS must be proposed based on the general structure of experimental data but not their values.
		\item[(3)] A common flavor structure not only for quarks but also for leptons without discrimination. Although the mixing angles of CKM and PMNS are different in numerical characterization, both are generated in charged current weak interactions and have the same parameterization. The similarity encourages us to build a general description of both quark flavors and lepton flavors.
	\end{itemize}

In this paper, we will address two major questions currently facing flavor physics: why SM fermions exhibit a mass hierarchical structure and what determines the flavor mixings of quarks and leptons.

We will start from the SM Yukawa couplings that generate quark and lepton mass matrices. After investigating the double roles of the diagonalizing transformation of the mass matrix in the mass spectrum and flavor mixing in Sec. \ref{sec.SMflavor}, we proposed a factorized decomposition of the mass matrix in Sec. \ref{sec.MFS}, which leads to the establishment of MFS.
The physical means of the factorized matrix are studies, which define a new Yukawa basis that is adopted to show an organized structure of fermion mass matrices.
Through the mass hierarchy limit condition, a $[SO(2)_L\times SO(2)_R]^f$ symmetry is found for the mass matrix of quarks/leptons. 
The continuous symmetry can be used to develop new flavor structure. 
Thus, mass hierarchy and flavor mixing can be clearly separated into two independent questions.
To parameterize the MFS, a close-to-flat mass structure is implied as the result of the homology hypothesis of fermion mass matrices.
In Sec. \ref{sec.fit}, the MFS is fit to lepton and quark masses, and CKM and PMNS experimental data are used to determine the generality and validity of our flavor structure
. Through the fit result and the means of the Yukawa phase, a flavor mixing sum rule is given in the hierarchy limit in Sec. \ref{sec.sumrule}, which further confirms the interpretation of the smallness of $s_{13}$.
In Sec. \ref{sec.unifiedYukawa}, we update an earlier work in \cite{ZhangMPLA2021} and generalize the close-to flat mass structure to quarks and leptons. In the end, a unified Yukawa interaction with universal couplings for all SM fermions is built.
In this paper, all fermions are SM fermions, adding minimal extended three Dirac neutrinos. The neutrino mass is treated as a normal order.

\section{Flavor Structure in the SM}\label{sec.SMflavor}
Regarding quantum numbers in the SM, elementary fermions are divided into up-type quarks $u$, down-type quarks $d$, neutrinos $\nu$ and charged leptons $e$. Each kind of fermion is organized in a 3-fold family structure.
Regarding gauge interactions, three fermionic families are not distinguished, i.e., gauge interactions have a family-universal form.
In phenomenology, they only differ by their mass and flavor mixing in charged current weak interactions. The latter is expressed by the Cabibbo-Kobayashi-Maskawa (CKM) mixing matrix for quarks and the Pontecorvo-Maki-Nakagawa-Sakata (PMNS) mixing matrix for leptons.
In the SM with minimal extended Dirac neutrinos, quark and lepton masses and flavor mixings all originate from Yukawa interactions.

The quark Yukawa terms are written as 
	\begin{eqnarray*}
		\mathcal{L}_{Y}^q=-y_{ij}^d\bar{Q}^i_LHd^j_R-y_{ij}^u\bar{Q}^i_L\tilde{H} u^j_R+H.c. ~~ (Q, u_R, d_R,\neq \psi_i^{u,d})
	\end{eqnarray*}
where $Q_L=\Array{c}{u_L \\ d_L}$ represents the left-handed quark doublet, $u_R,d_R$ represents the right-handed singlet, and $H$ represents a complex Higgs doublet scalar.
Yukawa interactions between different flavors are generated from flavor-dependent couplings $y_{ij}^u$ and $y_{ij}^d$ with the flavor index $i,j$. 
This is only flavor-dependent terms in the SM Lagrangian before electroweak symmetry spontaneously breaking (EWSB).
After EWSB, Higgs obtains its VEV $\langle H\rangle=v_0/\sqrt{2}$ and quarks become massive. The mass matrix can be expressed as 
$${M}_{ij}^q=\frac{v_0}{\sqrt{2}}{y}_{ij}^q,~~q=u,d.$$ 
Making chiral transformation in 3-fold family space as
	\begin{eqnarray}
		u_{Li}=({U}^u_{L})^\dag_{ij} u^{(m)}_{Lj},~
		u_{Ri}=({U}^u_{R})^\dag_{ij} u^{(m)}_{Rj},~
		d_{Li}=({U}^d_{L})^\dag_{ij} d^{(m)}_{Lj},~
		d_{Ri}=({U}^d_{R})^\dag_{ij} d^{(m)}_{Rj},
		\label{Eq.diagonalizetransf}
	\end{eqnarray}

${M}^q$ can be diagonalized to the mass basis labeled by superscript $^{(m)}$ under the biunitary transformation
\begin{eqnarray}
{M}^q\rightarrow {U}_{L}^q{{M}^q}({U}_R^q)^\dag={\rm diag}(m^q_1,m^q_2,m^q_3)
\label{Eq.diagM}
\end{eqnarray}
Here, $m^q_i$ is the $i$-th family quark mass.
Due to $U_L^u\neq U_L^d$, the diagonalizing transformations of ${M}^u$ and ${M}^d$ involve the flavor change effect in the weakly charged current interaction
	\begin{eqnarray}
		\mathcal{L}^q_{CC}
		=\frac{g}{\sqrt{2}}\bar{u}^{(m)}_L\gamma^\mu U_L^u (U_L^d)^\dag d^{(m)}_L W_\mu^+
			+H.c.+\cdots
	\end{eqnarray}
The CKM mixing matrix is defined by $U_{CKM}={U}^u_L{{U}^d_L}^\dag$.
Notably, the quark fields on a mass basis can always be redefined by a flavor-dependent phase such as
	\begin{eqnarray}
		u_{Li}\rightarrow e^{i\theta^u_i}u_{Li},~
		u_{Ri}\rightarrow e^{i\theta^u_i}u_{Ri},~
		d_{Li}\rightarrow e^{i\theta^d_i}d_{Li},~
		d_{Ri}\rightarrow e^{i\theta^d_i}d_{Ri}
	\end{eqnarray}
which maintains an invariant Lagrangian.

This means that some free parameters in $U_{CKM}$ can be eliminated by redefining quark fields. The process is called rephasing.
By rephasing six left-handed quark fields, five nonphysical phases in ${U}^u_L{{U}^d_L}^\dag$ can be eliminated.
The remaining four physical quantities in the CKM matrix are parameterized in the standard form with 3 mixing angles and 1 CP violation phase \cite{ClauPRL1984} as follows:
\begin{eqnarray*}
{U}_{CKM}=\left(\begin{array}{ccc}
		1&0&0\\
		0& c_{23} & s_{23}\\
		0& -s_{23} & c_{23}
		\end{array}\right)
		\left(\begin{array}{ccc}
		c_{13}&0 & s_{13}e^{-i\delta_{CP}}\\
		0 &1 &0\\
		-s_{13}e^{i\delta_{CP}} &0 & c_{13}
		\end{array}\right)
		\left(\begin{array}{ccc}
		c_{12} & s_{12} &0 \\
		-s_{12} & c_{12} &0\\
		0&0&1
		\end{array}\right)
\end{eqnarray*}
Here, $c_{ij}\equiv\cos\theta_{ij},s_{ij}\equiv \sin\theta_{ij}$. 
Three mixing angles can be calculated from 
	\begin{eqnarray*}
	 	s^2_{12}&=&\frac{|U_{CKM,12}|^2}{1-|U_{CKM,13}|^2},
		\\
		s^2_{23}&=&\frac{|U_{CKM,23}|^2}{1-|U_{CKM,13}|^2},
		\\
		s_{13}&=&|U_{CKM,13}|
	\end{eqnarray*}
The CPV can be determined by the Jarlskog invariant \cite{PetcovNPB2016}
	\begin{eqnarray*}
		J_{CP}&=&{\rm Im}\left(U_{CKM,23}U_{CKM,13}^*U_{CKM,12}U_{CKM,22}^*\right)
		\\
		&=&\frac{1}{8}\cos\theta_{13}\sin(2\theta_{12})\sin)(2\theta_{23})\sin(2\theta_{13})\sin\delta_{CP}
	\end{eqnarray*}
or equivalently from $|U_{CKM,22}|^2$ in  \cite{ZhangNPB2022}
	\begin{eqnarray*}
		|U_{CKM,22}|^2=c_{12}^2c_{23}^2+s_{13}^2s_{23}^2s_{12}^2-2c_{12}s_{12}c_{23}s_{23}s_{13}\cos(\delta_{CP}).
	\end{eqnarray*}

Alternatively, the Wolfenstein parameterization is also used to exhibit the mixing angle hierarchy $s_{13}\ll s_{23}\ll s_{12}\ll 1$ with four new parameters $\lambda,A,\bar{\rho}, \bar{\eta}$ \cite{WolfensteinPRL1983}
	\begin{eqnarray}
		&&\lambda=s_{12}=\frac{|U_{CKM,12}|}{\sqrt{|U_{CKM,11}|^2+|U_{CKM,12}|^2}},
		\\
		&&
		A\lambda^2=s_{23}=\lambda\frac{|U_{CKM,23}|}{|U_{CKM,12}|},
		\\
		&&s_{13}e^{i\delta_{CP}}=\frac{A\lambda^3(\bar{\rho}+i\bar{\eta})\sqrt{1-A^2\lambda^4}}{\sqrt{1-\lambda^2}[1-A^2\lambda^4(\bar{\rho}+i\bar{\eta})]}
	\end{eqnarray}
In phenomenology, Yukawa coupling $y_{ij}^q$ can be observed in phenomenology from 6 quark masses and 4 CKM mixing parameters.
These data were measured, and the results are listed in Tab. \ref{tab.quarkleptonexpdata}.
\begin{table}[htp]
\begin{center}
\caption{Physical masses and mixing parameters of quarks and leptons \cite{PDG2018}}
\begin{tabular}{|c|c|}
\hline
\hline
quark mass & CKM
\\
\hline
$\begin{array}{ll}
	m_u=2.2^{+0.5}_{-0.4}~{\rm MeV}
	&
	m_d=4.7^{+0.5}_{-0.3}~{\rm MeV}
	\\
	m_c= 1.275^{+0.025}_{-0.035}~{\rm GeV}
	&
	m_s=95^{+9}_{-3}~{\rm MeV}
	\\
	m_t= 173.0\pm0.4~{\rm GeV}
	&
	m_b=4.18^{+0.04}_{-0.03}~{\rm GeV}
	\end{array}$
	& 
	$\begin{array}{l}
	s_{12}=0.2244\pm0.0005
	\\
	s_{23}=0.0422\pm0.0008
	\\
	s_{13}=0.00394\pm0.00036
	\\
	\delta=(73.5^{+4.2}_{-5.1})^\circ
	\end{array}$
	\\
\hline
\hline
 lepton mass & PMNS
 \\
 \hline
	$\begin{array}{ll}
	m_e=0.5109989461(31)~{\rm MeV}
	&
	m_1^\nu=0.0001~{\rm eV (input)}
	\\
	m_\mu= 105.6583745(24)~{\rm MeV}
	&
	m_2^\nu=0.0086~{\rm eV}
	\\
	m_\tau= 1776.86(12)~{\rm MeV}
	&
    	m_3^\nu=0.050~{\rm eV}
	\end{array}$
	&
	$\begin{array}{l}
	s_{12}^2=0.297,~~0.250-0.354
	\\
	s_{23}^2=0.425,~~0.381-0.615
	\\
	s_{13}^2=0.0215(NH),~0.0190-0.0240
	\\
	\delta=1.38\pi,~~2\sigma: (1.0-1.9)\pi
	\end{array}$
\\
\hline\hline
\end{tabular}
\end{center}
\label{tab.quarkleptonexpdata}
\end{table}%

Now, we consider the lepton sector.
Considering the minimal extended SM with Dirac neutrinos, the lepton Yukawa terms are written as
	\begin{eqnarray*}
		\mathcal{L}_{Y}^l=-y_{ij}^e\bar{L}^i_L\tilde{H}e^j_R-y_{ij}^\nu\bar{L}^i_LH e^u_L+H.c. ~~ (L, e_R, \nu_R,\neq \psi_i^{e,\nu})
	\end{eqnarray*}
where $L_L=\Array{c}{\nu_L \\ e_L}$ represents the left-handed lepton doublet, and $\nu_R,e_R$ represents the right-handed singlet.
After EWSB, leptons become massive fields with the following mass matrix
	\begin{eqnarray*}
		{M}_{ij}^l=\frac{v_0}{\sqrt{2}}{y}_{ij}^l,~~l=\nu,e.
	\end{eqnarray*}
Physical masses can be obtained from biunitary diagonalizing transformation of ${M}^l$
	\begin{eqnarray}
		{M}^l\rightarrow {U}_{L}^l{{M}^l}({U}_R^l)^\dag={\rm diag}(m^l_1,m^l_2,m^l_3)
		\label{Eq.diagM}
	\end{eqnarray}
with the following unitary transformation
	\begin{eqnarray*}
		\nu_{Li}=({U}^\nu_{L})^\dag_{ij} \nu^{(m)}_{Lj},~
		\nu_{Ri}=({U}^\nu_{R})^\dag_{ij} \nu^{(m)}_{Rj},~
		e_{Li}=({U}^e_{L})^\dag_{ij} e^{(m)}_{Lj},~
		e_{Ri}=({U}^e_{R})^\dag_{ij} e^{(m)}_{Rj},
	\end{eqnarray*}
These transformations also involve lepton flavor mixing in charged current weak interactions.
	\begin{eqnarray}
		\mathcal{L}^l_{CC}
			&=&\frac{g}{\sqrt{2}}\bar{e}^{(m)}_L\gamma^\mu U_L^e (U_L^\nu)^\dag \nu^{(m)}_L W_\mu^+
			+H.c.+\cdots
	\end{eqnarray}
Here,  ${U}^e_L{{U}^\nu_L}^\dag$ is the PMNS mixing matrix $U_{PMNS}={U}^e_L{{U}^\nu_L}^\dag$.
Similarly, by using lepton field rephasing, the PMNS matrix can also be expressed by 3 mixing angles and 1 CPV.
Regarding the PMNS data, the lepton mixing parameters have two obviously larger angles ($\theta_{12}$ and $\theta_{23}$), and lepton mixing can produce the same flavor as that of quark mixing.

\section{The Minimal Flavor Structure}\label{sec.MFS}
The SM has not provided any values or structures of Yukawa couplings. As general complex quantities, these couplings include too many redundant d.o.f. To recover the fermion flavor structure, a faithful description of the mass matrix is first obtained with the same number of parameters as those of the physical observables. Then, a possible organization structure is determined.

In this section, we build the MFS in terms of two hypotheses. For the sake of convenience, we express all formulas in the quark sector, which can be directly generalized to the lepton sector.

\subsection{Yukawa Basis}
Considering the redefined chiral quark field by a free phase on a gauge basis obtains
	\begin{eqnarray*}
		\psi^q_{i,L,R}\rightarrow e^{i\theta^q_{i,L,R}}\psi^q_{i,L,R}
	\end{eqnarray*}
with $q=u,d$ for up-type and down-type quarks and family index $i=1,2,3$,
all the SM terms remain invariant except the Yukawa interaction and charged current weak interaction, which are just two terms related to the flavor problems.

In the Yukawa interaction, these complexes $\theta^q_{i,L,R}$ provide nontrivial phases between different families

	\begin{eqnarray*}
		\bar{\psi}_i^q\psi_j^{q'}\phi\rightarrow e^{i(\theta^{q'}_j-\theta^{q}_i)}\bar{\psi}_i^q\psi_j^{q'}\phi
	\end{eqnarray*}
which implies the origin of the CP violation.
This inspires us to propose the following hypothesis:
\\
{\bf Hypothesis I:} Complex phases in Yukawa couplings completely originate from the redefinition of fermion fields.
	
Because the Yukawa interaction is not determined by the gauge principle, this hypothesis indicates that the fermion mass matrix can be expressed as a real matrix on an appropriately chosen basis. We call it the Yukawa basis and label it by the superscript $^{(Y)}$ \cite{ZhangIJMPA2021}.
 
The quarks in the Yukawa basis and in the gauge basis have the following relation
 	\begin{eqnarray*}
		u_{L,R}^{(Y)}=F^u_{L,R}u_{L,R},~~~~
		d_{L,R}^{(Y)}=F^d_{L,R}u_{L,R},
	\end{eqnarray*}
Here, quarks are denoted as family triplets. The simplest rephasing transformation $F_{L,R}^{q}$ for $q=u,d$ is a diagonal matrix
	\begin{eqnarray*}
		F_{L}^{q}&=&{\rm diag}(e^{i\lambda_{L0}^q},e^{i\lambda_{L1}^q},e^{i\lambda_{L2}^q}),
		\\
		F_{R}^{q}&=&{\rm diag}(e^{i\lambda_{R0}^q},e^{i\lambda_{R1}^q},e^{i\lambda_{R2}^q}).
	\end{eqnarray*}
After EWSB, quark mass matrix $M_0^q$ is  $3\times 3$ real matrix
	\begin{eqnarray*}
		\mathcal{L}_M^q= -\bar{u}_{L}^{(Y)} M_0^u u_R^{(Y)}-\bar{d}_{L}^{(Y)} M_0^d d_R^{(Y)}+H.c.
	\end{eqnarray*}
Physical masses can be obtained by making a $SO(3)$ rotation 
	\begin{eqnarray}
	&u^{(Y)}_L=(U_0^u)^Tu^{(m)}_L,
	~~~
	&u^{(Y)}_R=(U_0^u)^Tu^{(m)}_R,
	\\
	&d^{(Y)}_L=(U_0^d)^Td^{(m)}_L,
	~~~
	&d^{(Y)}_R=(U_0^d)^Td^{(m)}_R.
	\end{eqnarray}
to diagonalize  $M_0^q$
	\begin{eqnarray}
		U_0^qM_0^q(U_0^q)^T={\rm diag}(m_1^q, m_2^q, m_3^q)
	\label{Eq.UMUdiag00}
	\end{eqnarray}
On the other hand, quark rephasing moves some physical phases into the weak charge current interaction as follows:
	\begin{eqnarray}
		\mathcal{L}^q_{CC}
			&=&\frac{g}{\sqrt{2}}\bar{u}^{(m)}_L\gamma^\mu  U_0^u F_L^u (F_L^d)^\dag(U_0^d)^\dag d^{(m)}_L W_\mu^+
			+H.c.
	\end{eqnarray}
That is, $F_L^u (F_L^d)^\dag$ contributes to the CKM mixing matrix
	\begin{eqnarray}
	U_{CKM}=U_0^u F_L^u (F_L^d)^\dag(U_0^d)^\dag
	\label{Eq.rephsingCKM01}
	\end{eqnarray}

These complex phases in $F_{L,R}^{u,d}$ have three properties in $U_{CKM}$:
	\begin{itemize}
		\item[(1)] Only left-handed $F_L^{u,d}$ contribute to CKM mixing. In addition, right-handed $F_R^{u,d}$ makes no contribution to phenomenology.
We can take $F_R^{u}=F_R^{d}={diag}(1,1,1)$, i.e., $\lambda_{Ri}^{u,d}=0$.
		\item[(2)] Due to
$F_L^u (F_L^d)^\dag ={\rm diag}(e^{i(\lambda_{L0}^u-\lambda_{L0}^d)},e^{i(\lambda_{L1}^u-\lambda_{L1}^d)},e^{i(\lambda_{L2}^u-\lambda_{L2}^d)})$, $U_{CKM}$ receives the contributions only from phase difference $\lambda_i^u-\lambda_i^d$. Without the loss of generality, $\lambda_{iL}^{d}$ can be set to zero.
		\item[(3)] Considering rephasing, a total phase in $U_{CKM}$ can be eliminated. The phases can be redefined as
			\begin{eqnarray*}
			\lambda_{L0}^u&\equiv&\lambda_0^u
			\\
			\lambda_{L1}^u&\equiv&\lambda_1^u+\lambda_0^u
			\\
			\lambda_{L2}^u&\equiv&\lambda_2^u+\lambda_0^u,
			\end{eqnarray*}
The CKM mixing matrix can be rewritten as
	\begin{eqnarray}
		U_{CKM}={e^{i\lambda_0^u}}U_0^u{\rm diag}(1,e^{i\lambda_1^u},e^{i\lambda_2^u}) (U_0^d)^T
	\label{Eq.umixing00}
	\end{eqnarray}
	\end{itemize}
Here, the global phase $\lambda_0^u$ can be eliminated by rephasing.

To date, by choosing the Yukawa basis, all complex phases in the quark mixing matrix are provided from only two quantities $\lambda_1^u$ and $\lambda_2^u$. This suggests a new explanation for the origin of CP violations from the relative phases of quarks between the gauge basis and the Yukawa basis. The mass matrix structure can also be considered in a real space.

\subsection{Flat Mass Matrix}
\subsubsection{Chiral $[SO(2)_L\times SO(2)_R]^q$ Symmetry in Hierarchy Limit}
In Eq. (\ref{Eq.umixing00}), complex phases have been separated from the CKM matrix; however, the two real rotations $U_0^{u}$ and $U_0^d$ remain unknown. Without loss of generality, they can be decomposed into a product of three rotations around independent axions
 \begin{eqnarray}
{U}_0^q
	={R}_3(\theta_3^q){R}_2(\theta_2^q){R}_1(\theta_1^q)
	\label{Eq.UR3R2R1}
\end{eqnarray}
Here, ${R}_i$ are defined as
\begin{eqnarray*}
	{R}_1(\theta)=\Array{ccc}{1 & 0 & 0 \\
		0 & c& s\\
		0 & -s & c},
	~~~
	{R}_2(\theta)=\Array{ccc}{c & 0 & -s \\
		0 & 1 & 0 \\
		s & 0 & c},
	~~~
	{R}_3(\theta)=\Array{ccc}{c & s & 0\\
		-s & c &0 \\
		0 & 0 & 1}
\end{eqnarray*}
with $s=\sin\theta,c=\cos\theta$.
Thus, the $U_{CKM}$ in Eq. (\ref{Eq.umixing00}) include 8 parameters: 6 rotation angles $\theta_i^{u},\theta_i^{d}$ (for $i=1,2,3$) and 2 phases $\lambda_{1,2}^u$. However, the desired number of parameters in $U_{CKM}$ is 4, corresponding to 3 CKM mixing angles and 1 CP violation phase. The reduction of redundant parameters can be achieved by analyzing the roles of ${R}_i$ in the mass hierarchy limit.

The diagonalized quark mass matrix exhibits a hierarchical structure 
$$M_{diag}^q=m_3^q\Array{ccc}{h_{12}^qh_{23}^q && \\ & h_{23}^q & \\ && 1}$$
 with the hierarchy $h_{ij}^q\equiv m_i^q/m_j^q$. 
In the mass hierarchy limit, $h_{ij}^q\rightarrow 0$, ${M}_{diag}^q$ is normalized by a total mass $m_\Sigma^q\equiv m_1^q+m_2^q+m_3^q$.
Moreover, it has the following simple form:
	\begin{eqnarray}
		\frac{1}{m_{\Sigma}^q}M_{diag}^q=\Array{ccc}{0 && \\ & 0& \\ && 1}
	\label{Eq.normalMassMatrix}
	\end{eqnarray}
Obviously, Eq. (\ref{Eq.normalMassMatrix}) exhibits chiral $SO(2)_L^q\times SO(2)_R^q$ symmetry in real space, that is, quark fields are invariant under transformations
	\begin{eqnarray}
		\psi_{L}^{q, (m)}\rightarrow{R}^T_3(\theta_L^q)\psi_{L}^{q, (m)},~~
		\psi_{R}^{q, (m)}\rightarrow{R}^T_3(\theta_R^q)\psi_{R}^{q, (m)}
	\label{Eq.quarkmassbasistrans}
	\end{eqnarray}
With the help of the chiral symmetry, the rotation angle $\theta_{3}^q$ in $R_3$ of Eq. (\ref{Eq.UR3R2R1}) can be absorbed into the chiral rotation angles $\theta_{L,R}^q$ in Eq. (\ref{Eq.quarkmassbasistrans}). They have no contribution to the mass matrix in the hierarchy limit but provide two random parameters to the CKM mixing matrix.

\subsubsection{Mass Matrix Reconstruction}
The left two rotations $R_1$ and $R_2$ can be used to reconstruct the quark mass matrix in terms of Eq. (\ref{Eq.UMUdiag00})
	\begin{eqnarray}
		{M}_0^q=m_{\Sigma}^q{R}_1^T(\theta_1^q){R}_2^T(\theta_2^q)\Array{ccc}{0 && \\ & 0& \\ && 1}{R}_2(\theta_2^q){R}_1(\theta_1^q)
	\label{Eq.massreconstruction}
	\end{eqnarray}
On the other hand, $R_1(\theta_1^{q})$ and $R_2(\theta_2^{q})$ for $q=u,d$ must meet the requirement of the CKM matrix.
Substituting Eq. (\ref{Eq.UR3R2R1}) into Eq. (\ref{Eq.umixing00}), the CKM matrix becomes
	\begin{eqnarray}
	{U}_{CKM}= {e^{i\lambda_0^u}}{R}_3(\theta_3^u){R}_2(\theta_2^u){R}_1(\theta_1^u){\rm diag}\left(1,e^{i\lambda^u_1},e^{i\lambda^u_2}\right){R}_1^T(\theta_1^d){R}_2^T(\theta_2^d){R}_3^T(\theta_3^d)
	\label{Eq.CKMRRR00}
	\end{eqnarray}
Because $({K}_L^u)^\dag {U}_{CKM}{K}_L^d$ is rephased with
${K}^{u}_L={\rm diag}(1,e^{i\beta^{u}_2},e^{i\beta^u_3}),
{K}^{d}_L={\rm diag}(e^{i\beta_{1}^{d}},e^{i\beta^{d}_2},e^{i\beta^{d}_3})$ has the same CKM mixing results as ${U}_{CKM}$, and Eq. (\ref{Eq.CKMRRR00}) can generally be expressed by
	\begin{eqnarray}
		{R}_1^T(\theta_1^u){R}_2^T(\theta_2^u){R}_3^T(\theta_3^u)\Big[({K}_L^u)^\dag {U}_{CKM}{K}_L^d\Big]{R}_3(\theta_3^d){R}_2(\theta_2^d){R}_1(\theta_1^d)
		=e^{i\lambda_0^u}{\rm diag}(1, e^{i\lambda^u_1},e^{i\lambda_2^u})
		\label{Eq.solve6theta}
	\end{eqnarray}
The above equation provides a way to investigate the rotation angles $\theta_i^{u,d}$ in terms of the CKM experimental data listed in Tab. \ref{tab.quarkleptonexpdata}.
If the CKM experimental data are input, regarding a set of fixed rephasing matrices $K_{L}^{u,d}$, the six angles $\theta_{i}^{u,d}$ that diagonalize $U_{CKM}$ to the unitary eigenvalues as the right side of Eq. (\ref{Eq.solve6theta}) can be solved.
In \cite{Zhang2021arXiv}, all possible ${K}^{u,d}_L$ have been scanned, and some results regarding $\theta_{i}^{u,d}$ have been obtained.

Using $R_1(\theta_1^q)$ and $R_2(\theta_2^q)$ from these calculation results, the structure of the reconstructed mass matrix can be analyzed.
Physicists are always interested in a common mass matrix structure between up-type quarks and down-type quarks.
This encourages us to propose another hypothesis:
\\
{\bf Hypothesis II:} In the hierarchy limit, normalized mass matrices of up-type and down-type quarks have a homological structure in the Yukawa basis.

Because $\theta_{1,2}^q$ determined the normalized ${M}_0^q/m_\Sigma^q$ through Eq. (\ref{Eq.massreconstruction}), Hypothesis II essentially requires $\theta_1^u=\theta_1^d$ and $\theta_2^u=\theta_2^d$.
The question of seeking homological $M_0^u$ and $M_0^d$ is studied by defining a deviation degree between normalized up-type and down-type mass matrices as follows:
	\begin{eqnarray}
		d_M= \frac{\sum_{i,j}\Big|(\frac{{M}_0^u}{m_{\Sigma}^u}-\frac{{M}_0^d}{m_{\Sigma}^d})_{ij}\Big|^2}{\sum_{i,j}\Big|(\frac{{M}_0^u}{m_{\Sigma}^u}+\frac{{M}_0^d}{m_{\Sigma}^d})_{ij}\Big|^2}
	\label{Eq.chifunction}
	\end{eqnarray}
When $\theta_1^u=\theta_1^d$ and $\theta_2^u=\theta_2^d$, $d_M$ has a minimal value $d_M=0$, and $\frac{{M}_0^u}{m_{\Sigma}^u}$ and $\frac{{M}_0^d}{m_{\Sigma}^d}$ obtain a homological structure.

The results in \cite{Zhang2021arXiv} show that $d_M$ tends to zero by evolving its value toward the minima.
A good solution is listed as follows:
	\begin{eqnarray}
		\theta_1^u=-0.7836,~~~
		\theta_2^u=0.6110,~~~
		\theta_3^u=-0.9362,
		\\
		\theta_1^d=-0.7858,~~~
		\theta_2^d=0.6199,~~~
		\theta_3^d=-0.7098,
		\\
		\lambda_1^u=0.0354,~~~
		\lambda_2^u=0.1000
	\end{eqnarray}
which corresponds to $d_M=0.00004122$.
The reconstructed quark mass matrices in Eq. (\ref{Eq.massreconstruction}) are
	\begin{eqnarray}
		{M}_{0}^{u}&=&\frac{1}{3}m_\Sigma^u\Array{ccc}{0.9874&0.9950&0.9986\\ 0.9950&1.0027&1.0063\\0.9986&1.0063&1.0099},~
		\\
		{M}_{0}^{d}&=&\frac{1}{3}m_\Sigma^d\Array{ccc}{1.0125&1.0035&1.0027\\ 1.0035&0.9945&0.9937\\ 1.0027&0.9937&0.9929}.
	\end{eqnarray}
The result strongly suggests that normalized mass matrices have a flat structure ${I}_0$ with all elements $1$
	\begin{eqnarray}
		\frac{1}{m_{\Sigma}^u}{M}_{0}^u=\frac{1}{m_{\Sigma}^d}{M}_{0}^d=\frac{1}{3}{I}_0,~~{I}_0\equiv\Array{ccc}{1&1&1\\1&1&1\\1&1&1}.
	\end{eqnarray}
The flat mass structure is a logical result from the homology requirement for the up-type and down-type quark mass matrices.
Historically, a similar democratic matrix for neutrino masses was assumed to explain the hierarchal structure and neutrino mixing \cite{Fukuura1999PRD} and was even applied to the quark sector in \cite{Fritzsch2017CPC}.

In contrast to the democratic matrix that is an assumed texture, the flat matrix is obtained from the CKM experiment data in Eq. (\ref{Eq.solve6theta}) and the mass matrix construction in the hierarchy limit.
In addition, the democratic matrix was used for only neutrinos (down-type quarks) and fixed charged leptons (up-type quarks) on a (close-to) mass basis. However, up-type and down-type quarks are treated equally in the MFS. ${M}_0^u$ and ${M}_0^d$ exhibit a flat structure at the same time.
Because complex phases in the mass matrix have been separated under Hypothesis I, the flat matrix occurs in a real flavor space, which means that flavor breaking will be achieved by real parameters that are to be discussed in the next section. While democratic matrices occur in a complex space, they need complex parameterization to break flavors. It also involves the differences to determine the origin of the CP violation.

\subsubsection{CKM Mixing Structure in Hierarchy Limit}
In terms of the flat mass matrix and $[SO(2)_L\times SO(2)_R]^{u,d}$ symmetry, the CKM mixing matrix can be determined by only four parameters, which is a requirement of the MFS.

An orthogonal rotation $S_0$ is defined as
	\begin{eqnarray}
		{S}_0=\frac{1}{\sqrt{6}}\Array{ccc}{\sqrt{3}& 0&-\sqrt{3}\\ -1& 2 & -1 \\ \sqrt{2} & \sqrt{2} & \sqrt{2}}
		\label{Eq.S0}
	\end{eqnarray}
that diagonalize the flat matrix ${I}_0$ 
	\begin{eqnarray}
	{S}_0{I}_0{S}_0^T ={\rm diag}(0,0,3).
	\end{eqnarray}

With the help of $S_0$ and $[SO(2)_L\times SO(2)_R]^{u,d}$ symmetry in the hierarchy limit, diagonalizing the transformations $U_0^{u}$ and $U_0^{d}$ in Eq. (\ref{Eq.UMUdiag00}) can generally be expressed as
	\begin{eqnarray}
		U_0^u={R}_3(\theta_L^u){S}_0,~~~
		U_0^d={R}_3(\theta_L^d){S}_0
	\end{eqnarray}
The quark mixing matrix $U_{CKM}$ in Eq. (\ref{Eq.rephsingCKM01}) becomes
\begin{eqnarray}
		U_{CKM}={R}_3(\theta^u)S_0\Array{ccc}{1&& \\ & e^{i\lambda_1^u} & \\ && e^{i\lambda_2^u}} S_0^T{R}_3^T(\theta^d)
		\label{Eq.UckmS00}
	\end{eqnarray}
Here, the trivial global factor $e^{i\lambda_0^u}$ has been eliminated. Now, the CKM mixing matrix is dominated by two rotation angles and two free Yukawa phases.

The three CKM mixing angles can be determined by these parameters
	\begin{eqnarray}
	 	s_{13}^2&=&\frac{1}{18}\left[-\sqrt{3}c_u(1-c_2)+(1-2c_1+c_2)s_u\right]^2
			+\frac{1}{18}\left[\sqrt{3}c_us_2+(-2s_1+s_2)s_u\right]^2
			\label{Eq.s130}
	\\
		\frac{s_{23}^2}{1-s_{13}^2}&=&\frac{1}{18}\left[(1-2c_1+c_2)c_u+\sqrt{3}(1-c_2)s_u\right]^2
			+\frac{1}{18}\left[c_u(2s_1-s_2)+\sqrt{3}s_2s_u\right]^2
				\label{Eq.s230}
	\\
		\frac{s_{12}^2}{1-s_{13}^2}&=&\frac{1}{36}\left[s_d(3(1+c_2)c_u+\sqrt{3}(-1+c_2)s_u)+c_d(\sqrt{3}c_u(1-c_2)-(1+4c_1+c_2)s_u)\right]^2
			\nonumber\\
			&&+\frac{1}{36}\left[s_2s_d(3c_u+\sqrt{3}s_u)-c_d(\sqrt{3}c_us_2+(4s_1+s_2)s_u)\right]^2
			\label{Eq.s120}
	\end{eqnarray}
where $c_q=\cos\theta^q,s_q=\sin\theta^q$ for $q=u,d$ and $c_i=\cos\lambda_i^u,s_i=\sin\lambda_i^u$ for $i=1,2$.
The Jarlskog invariant can be calculated as follows:
	\begin{eqnarray}
		J_{CP}&=&\frac{1}{54}\Bigg\{\sqrt{3}\sin(\frac{\lambda_2}{2})\sin(2\theta^u+2\theta^d)\Big[\cos(\lambda_1-\frac{3\lambda_2}{2})+\cos(\lambda_1+\frac{\lambda_2}{2})-2\cos(\frac{\lambda_2}{2})\Big]
		\nonumber\\
		&&+\sin(\lambda_1-\frac{\lambda_2}{2})
			\Big[
			-4\cos(\lambda_1-\frac{\lambda_2}{2})\sin(2\theta^u)\sin(2\theta^d)
			\nonumber\\
			&&+\cos(\frac{3\lambda_2}{2})[2\cos(2\theta^u-2\theta^d)+\cos(2\theta^u+2\theta^d)]
			+3\sin(\frac{\lambda_2}{2})\sin(2\theta^u+2\theta^d)\Big]
		\Bigg\}
\		\label{Eq.Jcp0}
	\end{eqnarray}
In the mixing structure, the requirement of the nonredundancy parameters has been satisfied.
Eq. (\ref{Eq.UckmS00}) includes four parameters: two $SO(2)_{L}^{u,d}$ angles $\theta^{u,d}$ and two Yukawa phases $\lambda_{1,2}^u$.

Eq. (\ref{Eq.UckmS00}) provides a CKM mixing structure that is not related to the quark mass data. Eq. (\ref{Eq.UckmS00}) is derived from the hierarchy limit, which means that CKM mixing is independent of the mass hierarchy. This results in the order of $h_{ij}^q$. As perturbations, the hierarchies $h_{ij}^q$ will contribute neglectable corrections to the CKM mixing parameters, which will be discussed in the next section. In contrast to this mixing structure, some earlier work in \cite{FritzschNPB1979, FritzschPLB1986, Fritzsch1987PLB} attempted to explain CKM mixing from quark masses.
To explain the two large mixing angles in PMNS from lepton masses, a new mass matrix of lepton that is different from the form of quarks must be assumed \cite{FritzschPLB1990}.
If flavor mixings in the quark sector and lepton sector are regarded from a homological manifestation, there is a common flavor mixing structure both for the CKM mixing matrix and for the PMNS mixing matrix. As we will see, Eq. (\ref{Eq.UckmS00}) will be generalized to leptons in Sec. \ref{sec.GenerToLepton}, which makes the MFS a candidate for the desired flavor structure of quarks and leptons.

\subsection{Flavor Breaking}\label{sec.flavorbreaking}
\subsubsection{Close-to-flat Mass Matrix}
\begin{figure}[htbp]
\begin{center}
\includegraphics[height=0.25 \textheight]{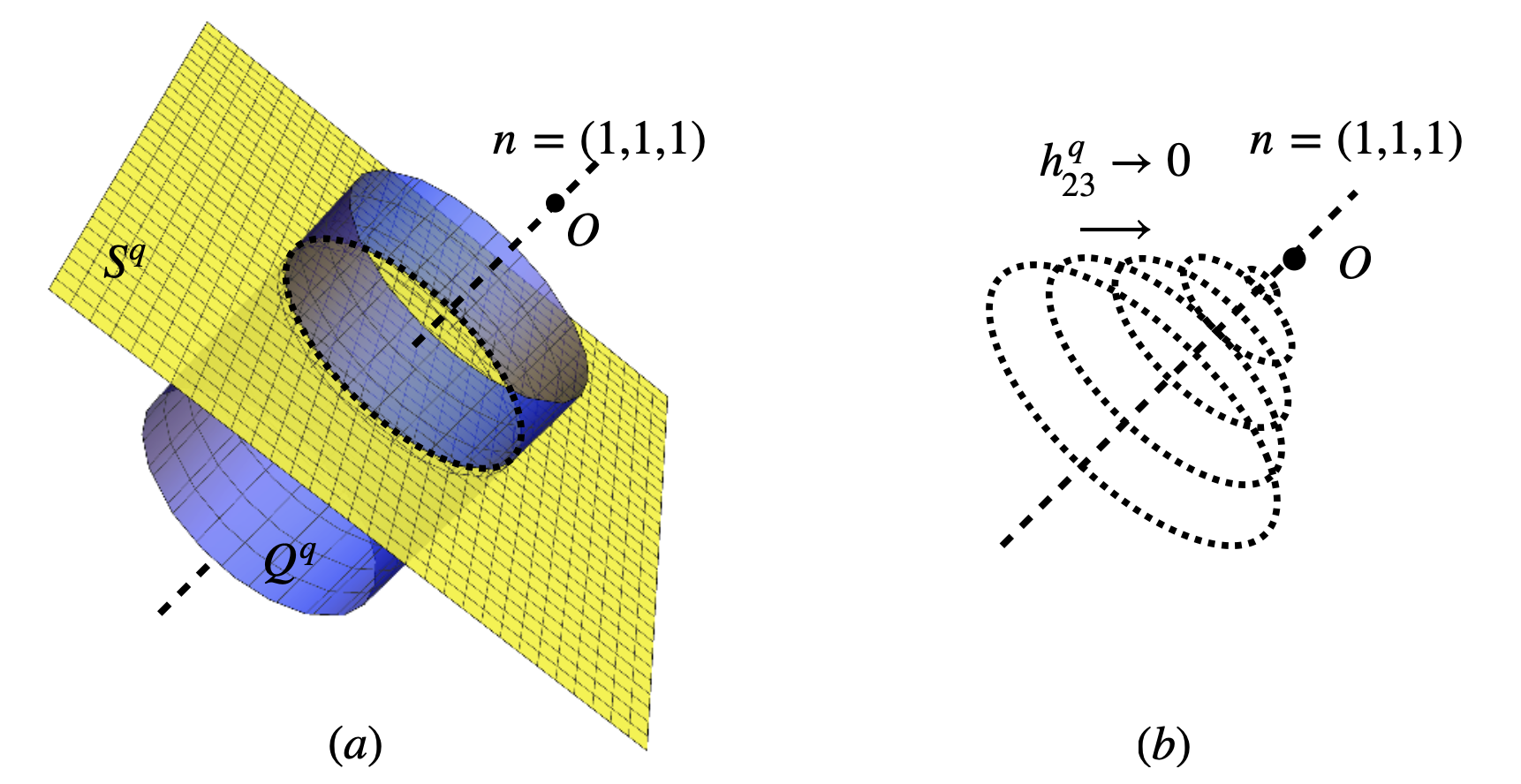}  
\caption{Mass Hierarchy and Breaking Flavor Parameters. (a) The intersection line (black dotted) of the surfaces $S^q$ (yellow) and cylindrical surface $Q^q$ (blue) forms a circle along $n=(1,1,1)$; (b) The circle collapses to the origin with $h_{23}^q\rightarrow 0$.}
\label{fig.explainVP}
\end{center}
\end{figure}

In mathematics, hierarchical eigenvalues can naturally arise from the close-to-flat matrix. We consider a $2\times2$ flat matrix with symmetric nondiagonal correction $\delta$
	\begin{eqnarray}
		\left(\begin{array}{cc}1& 1+\delta  \\ 1+\delta & 1 \end{array}\right).
	\end{eqnarray}
Its two eigenvalues $\chi_i$ are
	\begin{eqnarray}
		\chi_1&=&-\delta
		\\
		\chi_2&=&2+\delta
	\end{eqnarray}
A large hierarchy can be generated by the negative perturbation $\delta$
	\begin{eqnarray}
		\frac{\chi_1}{\chi_2}\simeq -\frac{\delta}{2}+\mathcal{O}(\delta^2).
	\end{eqnarray}
Inspired by the above property, we assume that the flat mass matrix is broken in a similar way.
We define a $3\times3$ matrix as follows
	\begin{eqnarray}
		{I}_\delta^q=\left(\begin{array}{ccc}1 & 1+\delta_{12}^q & 1+\delta_{13}^q\\ 1+\delta_{12}^q & 1 & 1+\delta_{23}^q \\ 1+\delta_{13}^q & 1+\delta_{23}^q & 1\end{array}\right)
	\end{eqnarray}
where $\delta_{ij}^q$ is the flavor-dependent nondiagonal correction.
The broken quark mass matrix ${M}_\delta^q$ can be written as
	\begin{eqnarray}
		{M}_\delta^q=\frac{m_{\Sigma}^q}{3}{I}_\delta^q
	\label{Eq.Mdelta}
	\end{eqnarray}
The small correction $\delta_{ij}^q$ breaks a flat ${I}_0$ to yield the two light flavors.
At a 1-order approximation of $\delta_{ij}^q$, the physical masses are dominated by the two parameters $S^q$ and $Q^q$ as follows
\begin{eqnarray}
		m^q_{1,2}&=&\frac{y^qv_0}{\sqrt{2}}\left(\frac{1}{3}S^q\mp\frac{2}{3}\sqrt{Q^q}\right)+\mathcal{O}(\delta^2)
		\label{Eq.m1m2}\\
		m^q_3&=&\frac{y^qv_0}{\sqrt{2}}\left(3-\frac{2}{3}S^q\right)+\mathcal{O}(\delta^2)
		 \label{Eq.m3}
	\end{eqnarray}
with 
	\begin{eqnarray}
		S^q&\equiv& -\delta^q_{12}-\delta^q_{23}-\delta^q_{13}
\label{Eq.S}\\
		Q^q&\equiv&(\delta^q_{12})^2+(\delta^q_{23})^2+(\delta^q_{13})^2-\delta^q_{12}\delta^q_{23}-\delta^q_{23}\delta^q_{13}-\delta^q_{13}\delta^q_{12}
\label{Eq.Q}
	\end{eqnarray}
In the space of $(\delta^q_{12},\delta_{23}^q,\delta_{13}^q)$, $S^q$ represents a plane with the normal direction $(1,1,1)$.
$Q^q$ represents a cylindrical surface along the axial direction $(1,1,1)$.
The intersection zone forms a circle, as shown in Fig. \ref{fig.explainVP} (a).
A random point on the circle meets the requirement of hierarchal masses.
From Eqs. (\ref{Eq.m1m2}) and (\ref{Eq.m3}), $S^q$ and $Q^q$ can be expressed as $h_{ij}^q$
	\begin{eqnarray}
		S^q&=&\frac{9(h^q_{23}+h^q_{12}h^q_{23})}{2(1+h^q_{23}+h^q_{12}h^q_{23})}
		\label{Eq.Spara0}\\
		&=&\frac{9}{2}(h^q_{23}+h^q_{12}h^q_{23}-(h^q_{23})^2)+\mathcal{O}(h^3)
		\nonumber\\
		Q^q&=&\frac{81}{4}\left((h^q_{23})^2
			-\frac{h^q_{23}(h^q_{23}+h^q_{12}h^q_{23})(1+2h^q_{23})}{(1+h^q_{23}+h^q_{12}h^q_{23})}
			+\frac{(1/2+h^q_{23})^2(h^q_{23}+h^q_{12}h^q_{23})^2}{(1+h^q_{23}+h^q_{12}h^q_{23})}
			\right)
		\label{Eq.Qpara0}\\
		&=&\frac{81}{16}(h^q_{23})^2+\mathcal{O}(h^3)
		\nonumber
	\end{eqnarray}
$h_{23}^q$ determines the position and radius of the circle in the leading order of $(h_{ij}^q)^2$. With $h_{23}^q$ tending to zero, the radius decreases to zero, and the circle moves to the origin along the axle $(1,1,1)$, as shown in \ref{fig.explainVP} (b).

\subsubsection{An Approximate $SO(2)_{LR}^q$ Symmetry}
Notably, the number of breaking flavor parameters is more than the requirement of the mass hierarchy.
To generate the two lighter quark masses, only two free parameters are needed to correspond to $h_{12}^q$ and $h_{23}^q$. As we have mentioned, $S^q,Q^q$ determines the hierarchy $h_{12}^q,h_{23}^q$ in Eqs. (\ref{Eq.Spara0}) and (\ref{Eq.Qpara0}). There are three perturbations $\delta_{ij}^q$ in the close-to-flat matrix $M_\delta^q$.
This means that there must exist a symmetry on three $\delta_{ij}^q$.
Fig. \ref{fig.explainVP} shows that each point $P$ on the circle corresponds to the same quark masses.

We define a distance between a point on the circle and the origin as follows
$$R^q=(\delta_{12}^q)^2+(\delta_{23}^q)^2+(\delta_{13}^q)^2,$$ is obviously invariant under the $SO(2)$ rotation along the direction of $(1,1,1)$ in the frame of $(\delta_{12}^q, \delta_{23}^q,\delta_{13}^q)$.
Choosing an initial point on the circle as $P(0)=-\frac{9}{4}h_{23}^q(1,1,0)$,
a random point $P(\theta)=\Big(\delta_{12}^q(\theta),\delta_{23}^q(\theta),\delta_{13}^q(\theta)\Big)$ can be expressed by the rotation angle $2\theta$
	\begin{eqnarray}
		\delta_{12}^q(\theta)&=&\Big(-\frac{3\cos(2\theta)}{4}+\frac{9\sin(2\theta)}{4\sqrt{3}}-\frac{3}{2}\Big)h_{23}+\mathcal{O}(h^2)
		\label{Eq.deltaTransdelta12}
		\\
		\delta_{23}^q(\theta)&=&\Big(-\frac{3\cos(2\theta)}{4}-\frac{9\sin(2\theta)}{4\sqrt{3}}-\frac{3}{2}\Big)h_{23}+\mathcal{O}(h^2)
		\label{Eq.deltaTransdelta23}
		\\
		\delta_{13}^q(\theta)&=&\Big(\frac{3\cos(2\theta)}{2}-\frac{3}{2}\Big)h_{23}+\mathcal{O}(h^2)
		\label{Eq.deltaTransdelta13}
	\end{eqnarray}
The mass matrix $M_\delta^q(\theta)$ generated from Point $P(\theta)$ has the same mass eigenvalues $(m_1^q,m_2^q,m_3^q)=(0,3h_{23},3-3h_{23})$ at the leading order of $h_{ij}^q$.
This indicates that the close-to-flat mass matrix $M_\delta^q$ must have a $SO(2)$ rotation symmetry.

We determine the $SO(2)$ representation on quark fields.
In the infinitesimal rotation of $P(\theta)$, $M_\delta^q$ transforms as follows
	\begin{eqnarray*}
		\delta M_\delta^q&=&\frac{9}{4\sqrt{3}}h_{23}^q\Array{ccc}{0 & 1 & 0 \\
			1 & 0 & -1 \\
			0 & -1 & 0 }\delta\theta.
	\end{eqnarray*}
Based on the $SO(2)_L\times SO(2)_R$ symmetry in the hierarchy limit, the symmetry of $M_\delta^q$ with nondiagonal corrections can be expressed into the chiral transformations generated by
	\begin{eqnarray}
		T_L&=&(1+a_Lh_{23})T_1+(1+b_Lh_{23})T_2+(1+c_Lh_{23})T_3
		\nonumber\\
		T_R&=&(1+a_Rh_{23})T_1+(1+b_Lh_{23})T_2+(1+c_Lh_{23})T_3
		\nonumber
	\end{eqnarray}
where $a_{L,R},b_{L,R}, and c_{L,R}$ are the perturbation corrections.
Here, $T_i$ for $i=1,2,3$ are $SO(2)$ generators.
	\begin{eqnarray}
		T_1=\Array{ccc}{0 & 0 & 0\\ 0 & 0 & -i \\ 0 & i & 0}
		,~~
		T_2=\Array{ccc}{0 & 0 & i\\ 0 & 0 & 0 \\ -i & 0 & 0}
		,~~
		T_3=\Array{ccc}{0 & -i & 0\\ i & 0 & 0 \\ 0 & 0 & 0}
	\nonumber
	\end{eqnarray}
which generate the rotation $R_i(\theta)=e^{iT_i\theta}$.
Regarding the vanishing perturbations $a_{L,R}=b_{L,R}= c_{L,R}=0$, it recovers to $SO(2)_L\times SO(2)_R$ symmetry.

Regarding an infinitesimal transformation, $T_{L,R}$ meets
\begin{eqnarray}
		iT_L^\dag M^q_{\delta}-iM^q_{\delta}T_R=\delta M^q_{\delta}+\mathcal{O}(h^2).
	\end{eqnarray}
The perturbation corrections are obtained as follows:
	\begin{eqnarray}
		&&a_L=0,~~b_L=-\frac{9}{4}, ~~c_L=0,
		\nonumber\\
		&&a_R=0,~~b_R=-\frac{9}{4},~~c_R=0
		\nonumber
	\end{eqnarray}
which stands for a $SO(2)_{LR}$ rotation along a shifted axle ${n}=(1,1-\frac{9}{4}h^q_{23},1)$, that is, $M^q_\theta$ is invariant under the rotation $R_{n}(\theta)$
	\begin{eqnarray}
		M^q_\delta(\theta)=R_{n}^T(\theta)M^q_\delta(0)R_{n}(\theta)
	\label{Eq.MthetaM0}
	\end{eqnarray}
Here, the $R_n(\theta)$ rotation along the random direction ${n}=(n_x,n_y,n_z)$ is defined by
	\begin{eqnarray}
		R_{n}(\theta)=\Array{ccc}{n_x^2(1-c)+c & n_xn_y(1-c)+n_zs & n_xn_z(1-c)-n_ys \\
			n_xn_y(1-c)-n_zs & n_y^2(1-c)+c & n_yn_z(1-c)+n_xs \\
			n_xn_z(1-c)+n_ys & n_yn_z(1-c)-n_xs & n_z^2(1-c)+c }
	\end{eqnarray}
where $c\equiv \cos\theta,s\equiv\sin\theta$, and ${n}$ is the normalized vector.

The above results mean that the chiral $[SO(2)_L\times SO(2)_R]^q$ symmetry of the flat mass matrix $M_0^q$ in the hierarchy limit is broken into an approximate $SO(2)_{LR}^q$ of $M_\delta^q$ in the order of $h_{ij}^q$ when flavor breaking is considered.

\subsubsection{Hierarchy Corrected Flavor Mixing}
In the hierarchy limit, two rotation angles in the CKM mixing matrix are provided by $SO(2)_L^u$ and $SO(2)_L^d$.
Regarding a close-to-flat mass matrix $M_\delta^q$, $SO(2)_L^u\times SO(2)_L^d$ is replaced by the approximate $SO(2)_{LR}^{u}\times SO(2)_{LR}^{d}$ symmetry to maintain the quark mixing structure.

We define a transformation $S_h^q$ with the $h_{23}^q$ correction to diagonalize $M^q_\delta(0)$
	\begin{eqnarray}
	 	\frac{1}{3m_\Sigma^q}S_h^qM_\delta^q(0)(S_h^q)^T={\rm diag}\Big(0,h_{23}^q,1-h_{23}^q\Big),
	\label{Eq.SMdeltaS}
	\end{eqnarray}
$S_h^q$ can be obtained as
	\begin{eqnarray*}
		S_h^q&=&S_0+\frac{h_{23}^q}{4\sqrt{3}}\Array{ccc}{
			0 & 0 & -0
			\\
			\sqrt{2} & \sqrt{2}  & \sqrt{2} 
			\\
			1 & -1 & 1}
			+\mathcal{O}\left((h_{23}^q)^2\right)
			\label{Eq.Scorrection00}
	\end{eqnarray*}
The above second term comes from the hierarchy corrections.

In terms of Eqs. (\ref{Eq.MthetaM0}) and (\ref{Eq.SMdeltaS}), $M_\delta^q(\theta)$ can be diagonalized by $S_h^qR_n^T(\theta)$
	\begin{eqnarray}
		\frac{1}{3m_\Sigma^q}\Big[S_h^qR_{n}(\theta)\Big]M_\delta^q(\theta)\Big[R_{n}^T(\theta)S^T\Big]
		=\frac{1}{3m_\Sigma^q}S_h^qM_\delta^q(0)(S_h^q)^T
		={\rm diag}(0,h_{23}^q,1-h_{23}^q)
	\label{Eq.MdeltaDiag}
	\end{eqnarray}
i.e. diagonalizing transformation of $M_\delta^q(\theta)$ is $U_0^q=S_h^qR_{n}(\theta)$.
Therefore, the CKM matrix with the 1-order hierarchy correction becomes
	\begin{eqnarray}
		U_{CKM}=\Big[S_h^uR_{n}(\theta^u)\Big]\Big[{\rm diag}(1, e^{i\lambda_1^u},e^{i\lambda_2^u})\Big]\Big[R_{n}^T(\theta^d)(S_h^d)^T\Big]
	\label{Eq.ClostToFlatCKM01}
	\end{eqnarray}
This structure can also be expressed in another form.
	\begin{eqnarray}
		U_{CKM}
		=\Big[R_{n_s}(\theta^u)S_h^u\Big]\Big[{\rm diag}(1, e^{i\lambda_1^u},e^{i\lambda_2^u})\Big]\Big[(S_h^d)^TR^T_{n_s}(\theta^d)\Big]
	\label{Eq.ClostToFlatCKM02}
	\end{eqnarray}
where $R_{n_s}(\theta^q)=S_h^qR_{n}(\theta^q)(S_h^q)^T$ is the definition.
Obviously, $R_{n_s}(\theta^q)$ is $R_3(\theta^q)$ corrected by the mass hierarchy.
It is easy to see that the above $U_{CKM}$ recovers to the result in the hierarchy limit in Eq. (\ref{Eq.UckmS00}) when $h_{23}^q\rightarrow 0$.

The hierarchy contributions to the CKM mixing angles can be expressed by
	\begin{eqnarray}
		\Delta s_{13}^2&=&C_{13}^uh_{23}^u+C_{13}^dh_{23}^d+\mathcal{O}(h^2)
		\label{Eq.Deltas13h1}\\
		\Delta s_{12}^2&=&C_{12}^uh_{23}^u+C_{12}^dh_{23}^d+\mathcal{O}(h^2)
		\label{Eq.Deltas12h1}\\
		\Delta s_{23}^2&=&C_{23}^uh_{23}^u+C_{23}^dh_{23}^d+\mathcal{O}(h^2)
		\label{Eq.Deltas23h1}
	\end{eqnarray}
Corrections to the Jarlskog invariant can also be expressed as
	\begin{eqnarray*}
		\Delta J_{CP}=C_{J}^uh_{23}^u+C_{J}^dh_{23}^d+\mathcal{O}(h^2)
		\label{Eq.DeltaJcph1}
	\end{eqnarray*}
All coefficients $C_{ij}^{u,d}$ and $C_J^{u,d}$ are listed in Appendix A.
These formulas indicate that the hierarchy effect only provides a small correction to the results in the hierarchy limit; that is, CKM mixing is dominated by an independent structure on the quark masses. Any large mixing angle or CP violating phase does not arise from mass hierarchies but is provided from the approximate $SO(2)_{LR}^{u,d}$ rotation angles and Yukawa phases.

\subsection{Generalizing to Lepton Sector}\label{sec.GenerToLepton}

Leptons also have hierarchal masses and exhibit flavor mixing in charged current weak interactions. Inspired by these flavor characteristics similar to those of quarks, the MFS can be generalized to leptons.

Applying Hypothesis I to the lepton sector, the lepton mass terms can be expressed by the real mass matrix $M_0^{\nu,e}$
\begin{eqnarray*}
		\mathcal{L}_M^l=-\bar{\nu}_{L}^{(Y)} M_0^\nu \nu_R^{(Y)}-\bar{e}_{L}^{(Y)} M_0^e e_R^{(Y)}+H.c.
	\end{eqnarray*}
on the Yukawa basis
\begin{eqnarray*}
		\nu_{L,R}^{(Y)}=F^\nu_{L,R}\nu_{L,R},~~~~
		e_{L,R}^{(Y)}=F^e_{L,R}e_{L,R}.
	\end{eqnarray*}
Making $SO(3)$ the rotation gives
\begin{eqnarray*}
		\nu^{(Y)}=(U_0^\nu)^T\nu^{(m)},~~~
		e^{(Y)}=(U_0^e)^Te^{(m)},
	\end{eqnarray*}
$M_0^{\nu,e}$ is diagonalized into the physical masses. The weakly interacting leptonic charged current on a mass basis becomes
\begin{eqnarray}
		\mathcal{L}^l_{CC}
			&=&\frac{g}{\sqrt{2}}\bar{e}^{(m)}_L\gamma^\mu  U_0^e F_L^e (F_L^\nu)^\dag(U_0^\nu)^\dag \nu^{(m)}_L W_\mu^+
			+H.c.
	\end{eqnarray}
The PMNS mixing matrix has the following structure:
\begin{eqnarray}
	U_{PMNS}=U_0^e F_L^e (F_L^\nu)^\dag(U_0^\nu)^\dag
	\end{eqnarray}

Lepton Yukawa phases have the same properties as $F_{L,R}^{u,d}$ in $U_{CKM}$
	\begin{itemize}
		\item[(1)] Right-handed $F_R^{\nu,e}$ has no contribution to lepton masses or PMNS. We can take $F_R^{\nu}=F_R^{e}={diag}(1,1,1)$.
		\item[(2)] Complex phases in the PMNS are only provided from the difference $\lambda_i^e-\lambda_i^\nu$ (for $i=0,1,2$). Without a loss of generality, the charged lepton $\lambda_i^e$ can be set to zero.
		\item[(3)] Because a global phase can be eliminated by rephasing, we take $\lambda_0^\nu=0$. Thus, there are only two free phases in $U_{PMNS}$
	\begin{eqnarray}
		U_{PMNS}=U_0^e{\rm diag}(1,e^{-i\lambda_1^\nu},e^{-i\lambda_2^\nu}) (U_0^\nu)^T
		\label{Eq.UpmnsExpress}
	\end{eqnarray}
	\end{itemize}

Under Hypothesis II, the homology of $M_0^\nu$ and $M_0^e$ must also be investigated in the same way as that adopted in the quark sector.

Using the PMNS data listed in Tab. (\ref{tab.quarkleptonexpdata}), the value of the deviation degree $d_M$ displays a tendency to zero, which means that the mass matrices of neutrinos and charged leptons may be addressed from a common matrix. A set of evolution results is listed in \cite{Zhang2021arXiv}.

\begin{eqnarray}
		\theta_1^\nu={-0.7857},~~~
		\theta_2^\nu={0.6152},~~~
		\theta_3^\nu={-0.1852},
		\\
		\theta_1^e={-0.7850},~~~
		\theta_2^e={0.6153},~~~
		\theta_3^e={-0.8217},
		\\
		\lambda_1^\nu=-0.5805,~~~
		\lambda_2^\nu=-1.7587,
	\end{eqnarray}
Using $\theta_{1,2}^{\nu,e}$ in the above results, ${M}_{0}^{\nu,e}$ are reconstructed as
\begin{eqnarray}
		{M}_{0}^{\nu}=\frac{1}{3}m_{\Sigma}^\nu\Array{ccc}{0.9993&1.0002&0.9995\\ 1.0002&1.0011&1.0004\\ 0.9995&1.0004&0.9997},
		\\
		{M}_{0}^{e}=\frac{1}{3}m_{\Sigma}^e\Array{ccc}{0.9995&0.9995&1.0003\\ 0.9995&0.9994&1.0003\\1.0003&1.0003&1.0011}.
	\end{eqnarray}
The results also strongly suggest a flat matrix as a common mass structure of the charged lepton and neutrino mass matrices.
Now, the flat mass matrices as a result of Hypothesis II become a common characteristic for not only up-type and down-type quarks but also neutrinos and charged leptons. It hints that quark and lepton Yukawa interactions may be described in a unified fashion.

Now, neutrino and charged lepton mass matrices are assumed to be
\begin{eqnarray}
		\frac{3}{m_{\Sigma}^l}{M}_{0}^l={I}_0\equiv\Array{ccc}{1&1&1\\1&1&1\\1&1&1}, ~l=\nu,e
	\end{eqnarray}
in the hierarchy limit. The PMNS obtains a similar result as the CKM in Eq. (\ref{Eq.UckmS00})
\begin{eqnarray}
		U_{PMNS}={R}_3(\theta^e)S_0\Array{ccc}{1&& \\ & e^{-i\lambda_1^\nu} & \\ && e^{-i\lambda_2^\nu}} S_0^T{R}_3^T(\theta^\nu)
		\label{Eq.PMNSstructureH0}
	\end{eqnarray}

Considering lepton flavor breaking, the three nondiagonal real corrections $\delta_{12}^l,\delta_{23}^l,\delta_{13}^l$ (for $l=\nu,e$) can also be introduced into the flat mass matrices of neutrinos and charged leptons
	\begin{eqnarray}
		{M}_\delta^l=\frac{m_\Sigma^l}{3}{I}_\delta^l=\left(\begin{array}{ccc}1 & 1+\delta_{12}^l & 1+\delta_{13}^l\\ 1+\delta_{12}^l & 1 & 1+\delta_{23}^l \\ 1+\delta_{13}^l & 1+\delta_{23}^l & 1\end{array}\right)
		\label{Eq.Mlepexpress}
	\end{eqnarray}
The physical masses are expressed in Eqs. (\ref{Eq.m1m2}) and (\ref{Eq.m3}) after replacing the superscript $^q$ with $^l$. The corrected PMNS has a similar structure as Eq. (\ref{Eq.ClostToFlatCKM01})
	\begin{eqnarray}
		U_{PMNS}=\Big[S_h^eR_{n}(\theta^e)\Big]\Big[{\rm diag}(1, e^{-i\lambda_1^\nu},e^{-i\lambda_2^\nu})\Big]\Big[R_{n}^T(\theta^\nu)(S_h^\nu)^T\Big]
	\label{Eq.ClostToFlatPMNS01}
	\end{eqnarray}
or in another form
	\begin{eqnarray}
		U_{PMNS}
		=\Big[R_{n_s}(\theta^e)S_h^e\Big]\Big[{\rm diag}(1, e^{-i\lambda_1^\nu},e^{-i\lambda_2^\nu})\Big]\Big[(S_h^\nu)^TR_{n_s}^T(\theta^\nu)\Big].
	\label{Eq.ClostToFlatPMNS02}
	\end{eqnarray}
To date, the lepton flavor structure is similar to that of quarks.

\subsection{Family Universal Yukawa Interaction}
Under Hypotheses I and II, the MFS has realized that
\begin{itemize}
	\item[(1)] The complex phases in the SM Yukawa couplings have been separated. On the Yukawa basis, Yukawa interactions can be rewritten with a family universal coupling. After EWSB is performed, the quark/lepton mass matrix becomes a real, close-to-flat matrix. The origin of the CP violation can be explained by the unitary phases between the Yukawa basis and gauge basis.	
	\item[(2)] Flavor mixing and mass hierarchy have been divided into two independent issues.
In the mass hierarchy limit, quark/lepton flavor mixing is dominated by two Yukawa phases and two $SO(2)$ rotation angles. It still provides a nonvanishing CP violation for the CKM and PMNS. Hierarchy corrections only play perturbative roles in flavor mixing. The flavor mixing matrix with 1-order hierarchy corrections retains a similar structure as in the hierarchy limit.
	\item[(3)] A close-to-flat mass matrix naturally arises from the homology of the mass matrix. The mass matrices of all fermions, up-type and down-type quarks, charged leptons and Dirac neutrinos, have a common structure due to their hierarchal mass characteristics. A family universal Yukawa coupling appears as an inevitable product of the hierarchal fermion masses and the homology of the mass matrices. In phenomenology, coupling is determined by the total family mass $m_\Sigma^f$.
	\end{itemize}
In the mass hierarchy limit, the rewritten Yukawa interaction is summarized as
\begin{eqnarray}
		-\mathcal{L}_Y
		&=&y^u\left(\bar{Q}_{L,1}^{(Y)}+\bar{Q}_{L,2}^{(Y)}+\bar{Q}_{L,3}^{(Y)}\right)\tilde{H}\left(\bar{u}_{R,1}^{(Y)}+\bar{u}_{R,2}^{(Y)}+\bar{u}_{R,3}^{(Y)}\right)
			\nonumber\\
			&&+y^d\left(\bar{Q}_{L,1}^{(Y)}+\bar{Q}_{L,2}^{(Y)}+\bar{Q}_{L,3}^{(Y)}\right){H}\left(\bar{d}_{R,1}^{(Y)}+\bar{d}_{R,2}^{(Y)}+\bar{d}_{R,3}^{(Y)}\right)
			\nonumber\\
			&&+(u\rightarrow \nu, d\rightarrow e)
		\label{Eq.FamilyUniversalYukawaTerm}
	\end{eqnarray}
This describes an indiscriminate interaction between the different fermion families, i.e., exchange invariant exists between two random families.

After EWSB is performed, the mass matrices have a flat structure as follows:
\begin{eqnarray*}
		-\mathcal{L}_M=\frac{v_0}{3\sqrt{2}}y^u\bar{u}_L^{(Y)}{I}_0 u_R^{(Y)}
		+\frac{v_0}{3\sqrt{2}}y^d\bar{d}_L^{(Y)}{I}_0 d_R^{(Y)}
		+\frac{v_0}{3\sqrt{2}}y^\nu\bar{\nu}_L^{(Y)}{I}_0\nu_R^{(Y)}
		+\frac{v_0}{3\sqrt{2}}y^e\bar{e}_L^{(Y)}{I}_0 e_R^{(Y)}+H.c.
	\end{eqnarray*}
The flat matrix $I_0$ has a chiral $[SO(2)_L\times SO(2)_R]^f$ flavor symmetry.
When $I_0$ is broken by nondiagonal perturbation corrections $\delta_{ij}^f$, an approximate $SO(2)_{LR}$ symmetry exists in $\mathcal{O}(h_{ij}^1)$, which remains the quark/lepton mixing structure in Eqs. (\ref{Eq.ClostToFlatCKM02}) and (\ref{Eq.ClostToFlatPMNS02}).

The MFS introduces the Yukawa basis to express fermion interactions with Higgs, which maintains all invariant gauge interactions. In terms of the EW interaction of the quarks, the following is obtained:
\begin{eqnarray*}
		\mathcal{L}^q_{EW}&=&i\bar{Q}_L\gamma^\mu \Big(\partial_\mu -i\frac{g}{\sqrt{2}}(\tau^+W_\mu^++\tau^- W_\mu^-)+ieQA_\mu
		-i\frac{g}{c_W}(\frac{\tau_3}{2}-Qs_W^2)Z_\mu\Big)Q_L
		\\
		&&+i\bar{u}_R\gamma^\mu \Big(\partial_\mu+ieQA_\mu+i\frac{g}{c_W}Qs_W^2\Big)u_R+(u_R\leftrightarrow d_R)
	\end{eqnarray*}
The natural current remains flavor-diagonal after mass diagonalization is performed through Eq. (\ref{Eq.diagonalizetransf})
\begin{eqnarray*}
		\mathcal{L}^q_{EW}&=&\frac{g}{\sqrt{2}}\bar{u}_L\gamma^\mu U_{CKM}d_L W_\mu^++h.c.
		-eQ^u\bar{u}_L\gamma^\mu u_LA_\mu
		-eQ^d\bar{d}_L\gamma^\mu d_LA_\mu
		\\
		&&+\frac{g}{c_W}\bar{u}_L\gamma^\mu (\frac{1}{2}-Q^us_W^2)Z_\mu u_L
		+\frac{g}{c_W}\bar{d}_L\gamma^\mu (-\frac{1}{2}-Q^ds_W^2)Z_\mu d_L+\cdots
	\end{eqnarray*}
Therefore, the MFS does not produce FCNC at the tree level.

Breaking the flat mass structure dynamically is an essential open question. These are some possible mechanisms to achieve flavor breaking.
	\begin{itemize}
		\item[(1)] by vacuum structure. The multi-Higgs model can provide extra vacuum expected values beyond the single one in the SM. As a typical example, two Higgs fields in 2HDM can assign different VEV and different couplings to fermions \cite{AlvesEPJC2021,GrzadkowskiPRD2016}.
		\item[(2)] by the flavon model. Flavor-independent interactions are a characteristic of flavones. By setting proper couplings, the MFS can be achieved \cite{Bazzocchi2004PRD, PascoliJHEP2016, KoideIJMPA2017}. For example, by considering the four flavons assisted by one Higgs scenario, three flavons $\phi_i$ can be assigned as VEVs as follows:	
	\begin{eqnarray}
		\langle\phi_1\rangle=\Array{c}{1 \\ 0 \\ 0},~~
		\langle\phi_2\rangle=\Array{c}{0 \\ 1 \\ 0},~~
		\langle\phi_3\rangle=\Array{c}{0 \\ 0 \\ 1}.~~
	\end{eqnarray}
The other $\phi_0$ is assigned a family universal VEV
	\begin{eqnarray}
		\langle\phi_0\rangle=\Array{c}{1 \\ 1 \\ 1}
	\end{eqnarray}
Thus, the MFS can be obtained as follows:
	\begin{eqnarray*}
		-\mathcal{L}_Y^q&=&\frac{y^u}{3}\bar{Q}_L^{(Y)}\phi_0\tilde{H}\phi_0^T u_R^{(Y)}
		+\frac{y^d}{3}\bar{Q}_L^{(Y)}\phi_0{H}\phi_0^T d_R^{(Y)}
		\\
		&&+\sum_{i\neq j}\frac{y^u}{3}\bar{Q}_L^{(Y)}\tilde{H}\phi_i\delta_{ij}^u\phi_j^T u_R^{(Y)}
		+\sum_{i\neq j}\frac{y^d}{3}\bar{Q}_L^{(Y)}\tilde{H}\phi_i\delta_{ij}^d\phi_j^T d_R^{(Y)}
	\end{eqnarray*}
	\item[(3)] by radiative corrections. In the hierarchy limit, only the third fermions are massive. As a flavor breaking seed, the unbalanced mass assignment may induce nonvanishing radiative corrections between the third family and the second one, i.e., it contributes to $\delta_{23}$. The flat mass matrix is broken as	
	\begin{eqnarray*}
			\Array{ccc}{1&1&1\\ 1&1& 1+\delta_{23} \\ 1 & 1+\delta_{23} & 1}
		\end{eqnarray*}
The mechanism makes the second family obtain a nonvanishing mass, which breaks the $SO(2)$ symmetry in the first two families. Moreover, the different masses between the first two families may induce new radiative correction to the first family. Thus, the MFS mass matrix can be induced from the third one to the second one and then to the first one. Calculating and verifying the mechanism is our future study. The mechanism will provide an exciting prospect: the flavor structure is self-determined in the MFS.
	\end{itemize}

\section{Fit Experiment Data}\label{sec.fit}
\subsection{MFS Parameters}
We have established the MFS of quarks and leptons with the correct number of parameters. 
In the quark/lepton sector, some parameters are responsible for fermion masses, and others are responsible for flavor mixing.

\begin{figure}[htbp]
\begin{center}
\includegraphics[height=0.35 \textheight]{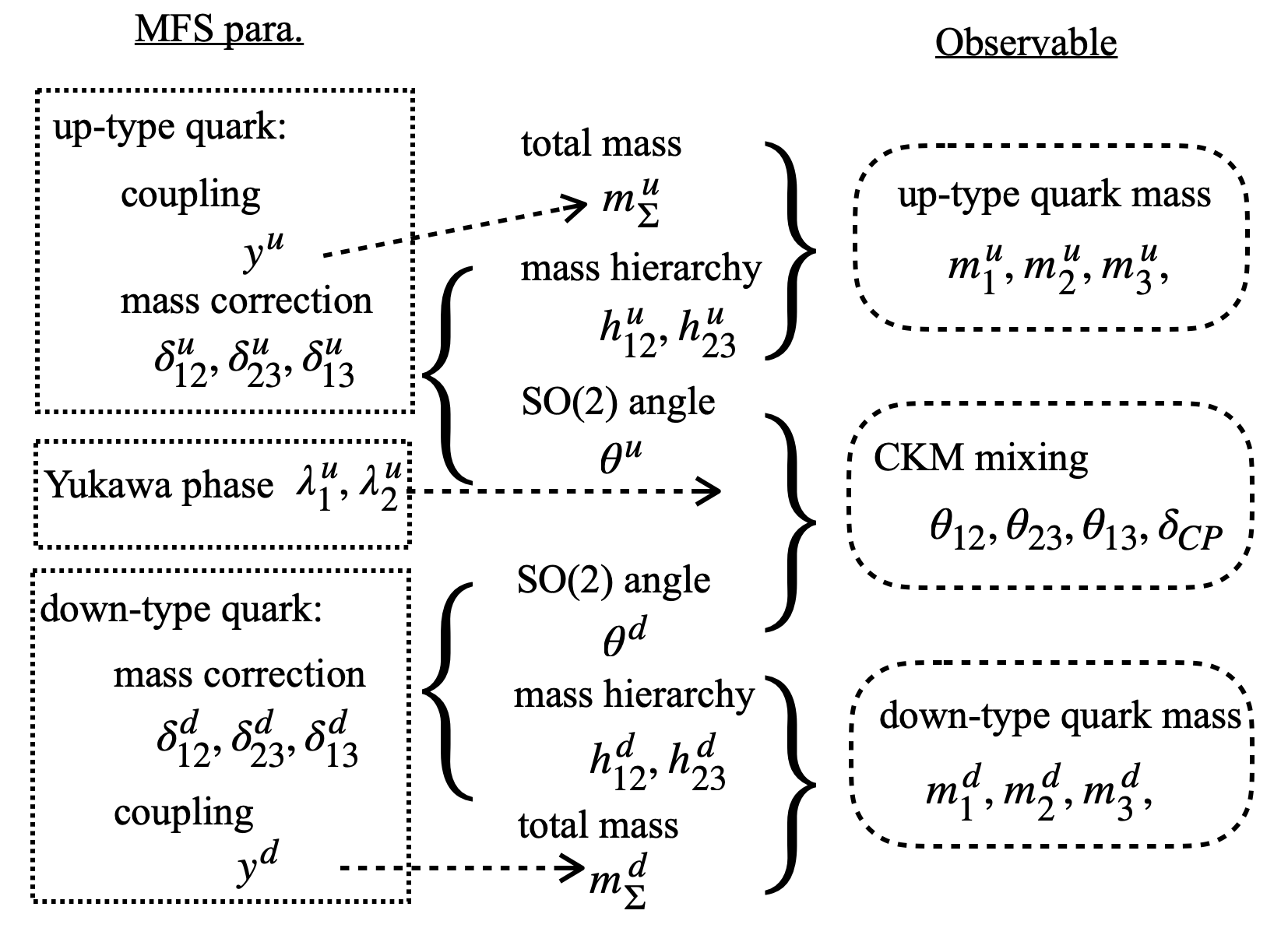}
\caption{MFS parameters and observables in the quark sector.}
\label{fig.MFSpara}
\end{center}
\end{figure}

On the Yukawa basis, the close-to-flat mass matrix has 3 nondiagonal real corrections $\delta_{12}^f,\delta_{23}^f,\delta_{13}^f$. They determine 2 mass hierarchies $h_{12}^f, h_{23}^f$ in terms of Eq. (\ref{Eq.m1m2}) and (\ref{Eq.m3}) and ignore one free $SO(2)_{LR}^f$ rotation angle.
Two rotation angles of $SO(2)_{LR}^{u}$ and $SO(2)_{LR}^{d}$ control the CKM flavor mixing with 2 Yukawa couplings together. (For leptons, the two rotation angles correspond to the $SO(2)_{LR}^{\nu}$ and $SO(2)_{LR}^e$ symmetries.) The total family mass $\sum_i m_i^f$ determines Yukawa couplings $y^f$. These MFS parameters are shown in Fig. \ref{fig.MFSpara}.
The MFS has provided 10 independent quantities to parameterize the quark (lepton) flavor structure: 2 Yukawa phases, 3+3 mass corrections $\delta_{ij}^f$ and 2 family universal Yukawa couplings. We will calculate these parameters by fitting fermion masses and flavor mixing parameters.

\subsection{Quark Masses and CKM}

To fit the MFS parameters, a two-step strategy is adopted. The first step is to fit 3 $\theta_{ij}^{u,d}$ to the quark masses. Using these fit results, the quark mass matrix ${M}_0^{u,d}$ can be reconstructed in terms of Eq. (\ref{Eq.Mdelta}).
The second step is to fit the CKM by calculating the diagonalizing transformations of ${M}_0^u$ and ${M}_0^d$.

The family universal Yukawa coupling can be determined by the total family mass
\begin{eqnarray*}
		\frac{v_0}{\sqrt{2}}y^q=\frac{1}{3}m_{\Sigma}^q
	\end{eqnarray*}
as
\begin{eqnarray*}
		y^u=0.334,~~~
		y^d=0.00820
	\end{eqnarray*}
where the VEV is $v_0=246$GeV, and the quark mass data is listed in Tab. (\ref{tab.quarkleptonexpdata}).

\begin{figure}[htbp]
\begin{center}
\includegraphics[height=0.22 \textheight]{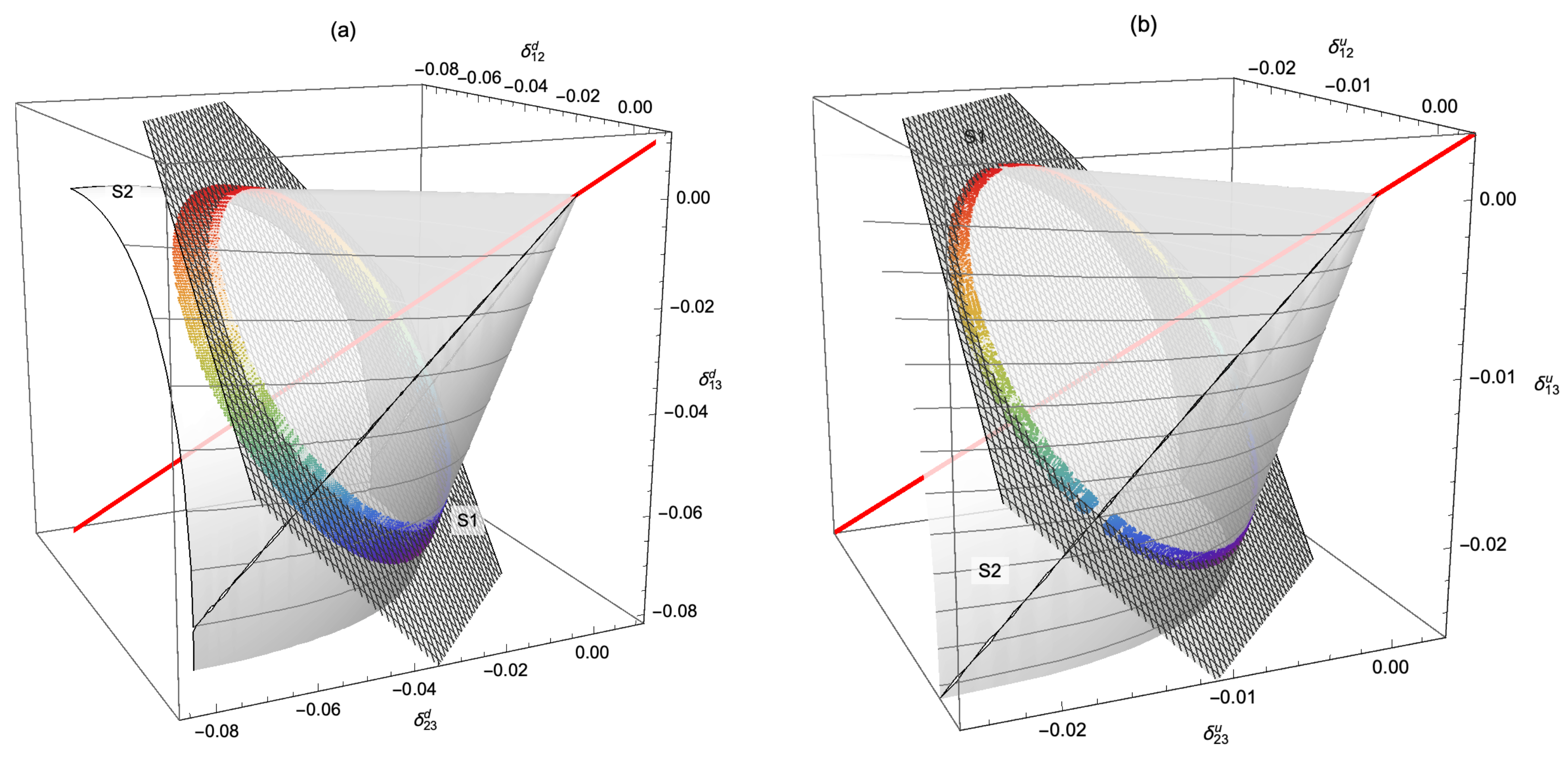}
\caption{Allowed parameter spaces $(\delta_{12}^q,\delta_{23}^q,\delta_{13}^q)$ for (a) down-type quarks and (b) up-type quarks in $1\sigma$ CL.}
\label{fig.mdmu}
\end{center}
\end{figure}

For a set of corrections $\delta_{ij}^q$ ($q=u,d$), the eigenvalues of ${M}_0^q$ can be calculated as physical quark masses.
By scanning all possible ranges of $\delta_{ij}^q$, a possible $\delta_{ij}^q$ is recorded when the eigenvalues of ${M}^q$ satisfy the quark mass data in $1\sigma$ C.L.
These fit values are shown in the frame of $(\delta_{12}^q,\delta_{23}^q,\delta_{13}^q)$ in Fig. \ref{fig.mdmu}, which forms a circle.
The fit results show that $SO(2)_{LR}^{q}$ is a good approximate symmetry, as we have analyzed before.
All points on a circle correspond to two hierarchies $h_{12}^q,h_{23}^q$.
The succession of the reproduction of the quark masses verify that the close-to-flat mass matrix is a good structure that addresses the quark mass hierarchy.

In terms of the above fitting results of $\delta_{ij}^{u}$ and $\delta_{ij}^{d}$, quark mixing can be investigated.
In Eq. (\ref{Eq.umixing00}), $U_{CKM}$ is determined by two rotation transformations $U_0^u, U_0^d$ and two Yukawa phases $\lambda_1^u,\lambda_2^u$.
By choosing a fit point $(\delta_{12}^u,\delta_{23}^u,\delta_{13}^u)$ on the up-type circle, we can build the up-type quark mass matrix through Eq. (\ref{Eq.Mdelta}). The $U_0^u$ transformation is calculated by diagonalizing the real ${M}_0^u$ into the physical mass, as shown in Eq. (\ref{Eq.UMUdiag00}). Similarly, for a down-type quark, $U_0^d$ can also be obtained from a fitting point $(\delta_{12}^d, \delta_{23}^d,\delta_{13}^d)$.
Thus, for two sets of chosen points $(\delta_{12}^u,\delta_{23}^u,\delta_{13}^u)$ and $(\delta_{12}^d,\delta_{23}^d,\delta_{13}^d)$, quark mixing is only dependent on the left two free parameters $\lambda_{1}^u,\lambda_{2}^u$.
By scanning the whole parameter space of $\lambda_{1}^u$ and $\lambda_{2}^u$, we can determine
the group of $(\delta_{12}^u,\delta_{23}^u,\delta_{13}^u)$ and $(\delta_{12}^d,\delta_{23}^d,\delta_{13}^d)$.
We investigate all possible combinations of $\delta_{ij}^u$ and $\delta_{ij}^d$ and obtain a set of good mixing results listed in the first column of Tab. \ref{Tab.CKMfit00}. The mass hierarchies $h_{ij}^q$ are also shown in the second column.
\begin{table}[htp]
\begin{center}
\caption{MFS fit results for quarks}
\begin{tabular}{c|c|c}
\hline
\hline
MFS Para. & Mass Hierarchy & CKM Mixing
\\
\hline
$\left\{\begin{array}{l}
	\delta_{12}^u=-0.0000453
	\\
	\delta_{23}^u=-0.0172
	\\
	\delta_{13}^u=-0.0165
	\end{array}\right.$
	&
	$\left.\begin{array}{l}
	\\
	h_{12}^u=0.00169
	\\
	h_{23}^u=0.00754
	\end{array}\right.$
	&$\left.\begin{array}{l}
		\\
		\\
		s_{12}=0.2243
	\end{array}\right.$

\\
	$\left.\begin{array}{l}
	\lambda_1^u=-0.00504
	\\
	\lambda_2^u=0.0851
		\end{array}\right.$
	& &
	$\left.\begin{array}{l}
		s_{23}=0.04141
		\\
		s_{13}=0.003942
	\end{array}\right.$

\\
$\left\{\begin{array}{l}
	\delta_{12}^d=-0.00723
	\\
	\delta_{23}^d=-0.0644
	\\
	\delta_{13}^d=-0.0377
	\end{array}\right.$
	&
	$\left.\begin{array}{l}
		h_{12}^d=0.0480
		\\
		h_{23}^d=0.0237
		\\
		\\
	\end{array}\right.$
	&
		$\left.\begin{array}{l}
		\delta_{CP}=1.31027
		\\
		\\
		\\
	\end{array}\right.$
	\\
\hline
\hline
\end{tabular}
\end{center}
\label{Tab.CKMfit00}
\end{table}%

The fit results for the magnitudes of the CKM matrix elements are
\begin{eqnarray*}
	U_{CKM}=\Array{ccc}{
0.975 & 0.224 & 0.00394
\\
 0.224 & 0.974 & 0.0414
 \\
 0.00909 & 0.0406 & 0.999
}
	\end{eqnarray*}
Alternatively, the fit CKM results may be expressed by the Wolfenstein parameters as follows:
\begin{eqnarray*}
	&\lambda=0.224, ~~~~
	&A=0.823,
	\\
	&\bar\rho= 0.107, ~~~~
	&\bar\eta=0.400.
	\end{eqnarray*}
These results agree well with the experimental measurements in \cite{PDG2018}. To date, all 10 parameters in the MFS have been determined by 6 quark masses, 3 CKM mixing angles and 1 CP violating phase.

Using the above fit results, the validity of the CKM mixing structure in $\mathcal{O}(h_{ij}^1)$ can also be verified by calculating the $SO(2)_{LR}^{u,d}$ rotation angles from the fit $\delta_{ij}^{u,d}$.
The rotation angle $\theta^{u,d}$ can be determined from Eq. (\ref{Eq.ClostToFlatCKM01}) as follows:
	\begin{eqnarray*}
		{R}_n(\theta^q)=(S_h^q)^\dag U_0^q
	\end{eqnarray*}
Substituting $\delta_{ij}^q$ listed in Tab. \ref{Tab.CKMfit00} to generate transformation $U_0^q$, the rotation angles can be obtained as follows:
	\begin{eqnarray*}		
		\theta^u=-1.029,~~~
		\theta^d=-0.8044.
	\end{eqnarray*}

Frm the above rotation angles $\theta^u,\theta^d$ and Yukawa phases $\lambda_1^u,\lambda_2^u$, the CKM in $\mathcal{O}(h^1)$ can be retrieved only with a slight difference from the experimental data. It is verified that $SO(2)^u\times SO(2)^d$ is a good approximate symmetry in the order of $h_{ij}^1$.

\subsection{Lepton Masses and PMNS}
Along the same two-step path, the MFS can be determined in the lepton sector.

\begin{figure}[htbp]
\begin{center}
\includegraphics[height=0.22 \textheight]{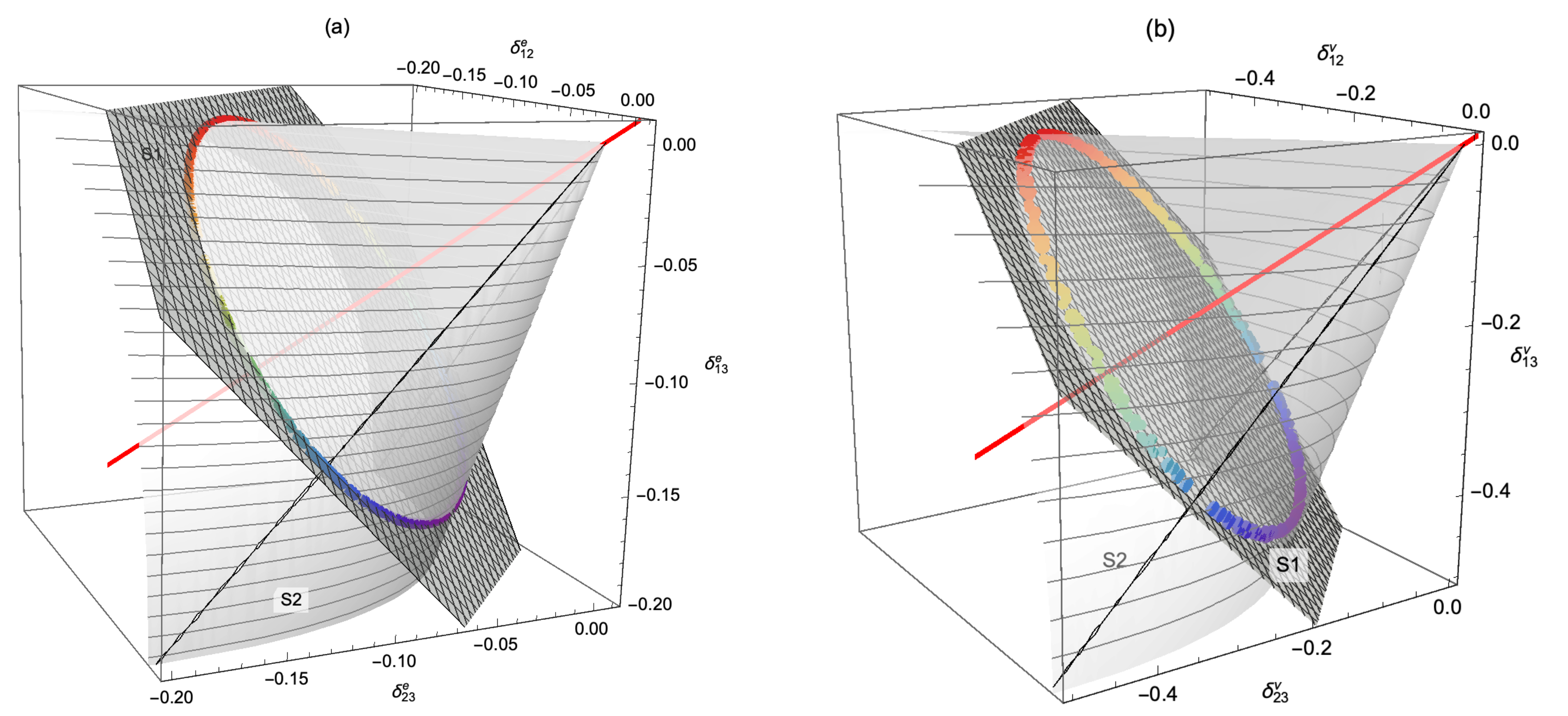}
\caption{Allowed parameter spaces $(\delta_{12}^l,\delta_{23}^l,\delta_{13}^l)$ for (a) charged leptons and (b) neutrinos.}
\label{fig.memnu}
\end{center}
\end{figure}

Due to the unknown absolute neutrino masses, the lightest neutrino mass $m_1^\nu$ is set to $0.0001$ eV. $m_2^\nu$ and $m_3^\nu$ are determined from $\Delta m^2_{32}$ and $\Delta m^2_{21}$ (for normal order).
In the first step, $\delta_{ij}^\nu$ and $\delta_{ij}^e$ are fit to the neutrino masses and charged lepton masses listed in Tab. \ref{tab.quarkleptonexpdata} by scanning the whole parameter space. All allowed $\delta_{ij}^{\nu,e}$ values are shown in Fig. (\ref{fig.memnu}) for neutrinos and charged leptons. These allowed distributions form the circles in the frame of $(\delta_{12}^l,\delta_{23}^l, \delta_{13}^l)$ for $l=\nu,e$ along the axle $(1,1,1)$.

By choosing two sets of $\delta_{ij}^{\nu}$ and $\delta_{ij}^e$ from the respective fit data, we can rebuild the neutrino and charged lepton mass matrices in terms of Eq. (\ref{Eq.Mlepexpress}). By calculating the transformations ${U}_0^\nu$ and ${U}_0^e$, we can obtain the PMNS mixing matrix through Eq. (\ref{Eq.UpmnsExpress}) with two free Yukawa phases $\lambda_{1,2}^\nu$. By scanning all possible combinations of the fit $\delta_{ij}^\nu$ and $\delta_{ij}^e$ and phases $\lambda_{1,2}^\nu$, we can obtain a fit result to match the PMNS experimental data listed in Tab. \ref{tab.PMNSfit01}.
The validity of the PMNS mixing structure can be verified by calculating the $SO(2)_{LR}^{\nu,e}$ rotation angles.
The two rotation angles are
\begin{eqnarray*}		
		\theta^\nu=-0.3837,~~~
		\theta^e=-0.9046,
	\end{eqnarray*}
which can generate a good PMNS mixing matrix in terms of Eq. (\ref{Eq.ClostToFlatPMNS02}).
\begin{table}[htp]
\begin{center}
\caption{MFS fit results for leptons}
\begin{tabular}{c|c|c}
\hline
\hline
MFS Para. & Mass Hierarchy & PMNS Mixing
\\
\hline
$\left\{\begin{array}{l}
	\delta_{12}^e=-0.00443
	\\
	\delta_{23}^e=-0.146
	\\
	\delta_{13}^e=-0.105
	\end{array}\right.$
	&
	$\left.\begin{array}{l}
	\\
	h_{12}^e=0.00484
	\\
	h_{23}^e=0.0594
	\end{array}\right.$
	&$\left.\begin{array}{l}
		\\
		\\
		s^2_{12}=0.335
	\end{array}\right.$

\\
	$\left.\begin{array}{l}
	\lambda_1^\nu=-0.0111
	\\
	\lambda_2^\nu=1.60
		\end{array}\right.$
	& &
	$\left.\begin{array}{l}
		s_{23}^2=0.439
		\\
		s_{13}^2=0.0201
	\end{array}\right.$

\\
$\left\{\begin{array}{l}
	\delta_{12}^\nu=-0.176
	\\
	\delta_{23}^\nu=-0.435
	\\
	\delta_{13}^\nu=-0.073
	\end{array}\right.$
	&
	$\left.\begin{array}{l}
		h_{12}^\nu=0.0116
		\\
		h_{23}^\nu=0.173
		\\
		\\
	\end{array}\right.$
	&
		$\left.\begin{array}{l}
		\delta_{CP}=1.49\pi
		\\
		\\
		\\
	\end{array}\right.$
	\\
\hline
\hline
\end{tabular}
\end{center}
\label{tab.PMNSfit01}
\end{table}%


\section{Flavor Mixing Sum Rule}\label{sec.sumrule}
The MFS provides a flavor mixing structure with four parameters in Eq. (\ref{Eq.UckmS00}) for quarks and Eq. (\ref{Eq.PMNSstructureH0}) for leptons.
In the CKM/PMNS matrix, the number of independent observables is also four. Therefore, there is no prediction regarding the flavor mixing parameters.
However, in the fit results listed in Tabs. \ref{Tab.CKMfit00} and \ref{tab.PMNSfit01}, both the quark $\lambda_1^u$ and lepton $\lambda_1^\nu$ have small fit values. This hints that $\lambda_1^{f}$ (for $f=u,\nu$ for quark and lepton) may be treated as a perturbation. This can be explained by the role played by the Yukawa phases. As we know, in the hierarchy limit, the physical mass becomes $(0,0,m_3^f)$. $\lambda_1^f,\lambda_2^f$ are two relative phases between three quark/lepton generations, which can be factorized into
\begin{eqnarray*}
	   \Array{ccc}{1 && \\ & e^{i\lambda_1^f} & \\ && e^{i\lambda_2^f}}
	  =\Array{cc}{\Array{cc}{1 & \\ & e^{i\lambda_1^f}} & \\ & 1}
	  \Array{cc}{\Array{cc}{1 & \\ & 1}&\\ &  e^{i\lambda_2^f}}
	\end{eqnarray*}
This means that the phase $\lambda_2^f$ in the second factorized matrix on the right side represents a phase between the 3rd generation and the whole of the first two generations.
The phase $\lambda_1^f$ parameterizes a phase between the 1st generation and 2nd generation. In the mass hierarchy limit, the first two generations become massless. The effect of the relative phase $\lambda_1^f$ is suppressed by hierarchy. In addition, $\lambda_2^f$ dominates complex phase roles in flavor mixing.
If treating $\lambda_1^f=0$, quark/lepton mixing structures are determined by only three independent parameters, which will yield a sum rule on the mixing angles and CPV.

Considering the CKM mixing matrix in Eq. (\ref{Eq.UckmS00}), we take $\lambda_1^u=0$, and we reduce Eq. (\ref{Eq.s120}) to
	\begin{eqnarray*}
		\frac{s_{12}^2}{1-s_{13}^2}&=&\frac{1}{36}\Big(\sin^2(\lambda^u_2)(\sqrt{3}c_{u+d}+c_ds_u-3c_us_d)^2\Big)
		\\
		&&+\frac{1}{36}\Big(\sqrt{3}(\cos(\lambda^u_2)-1)c_{u+d}+(5+\cos(\lambda^2_2))s_uc_d
		-3(1+\cos(\lambda^u_2))c_us_d\Big)^2
	\end{eqnarray*}
Moreover, the expression is rearranged in terms of $\theta^d$ as follows:
	\begin{eqnarray}
		C_{12}s_d^2+C_{1sc}s_dc_d+C_{10}=0
	\label{Eq.sumruleeq01}
	\end{eqnarray}
with the coefficients as follows:
	\begin{eqnarray*}
		C_{12}&=&\Big\{2(1-\cos(\lambda^u_2))(4\sqrt{3}s_uc_u+8s_u^2-12c_u^2)+36(c_u^2-s_u^2)\Big\}
		\\
		C_{1sc}&=&\Big\{2(1-\cos(\lambda^u_2))(4\sqrt{3}s^2_u+12s_uc_u)-72s_uc_u\Big\}
		\\
		C_{10}&=&\Big\{2(1-\cos(\lambda^u_2))(3-4\sqrt{3}s_uc_u-8s_u^2)+36s_u^2\Big\}-\frac{36s_{12}^2}{1-s_{13}^2}
		\\
	\end{eqnarray*}
Similarly, $|U_{CKM,22}|^2$ is rewritten in terms of $\theta^d$ as follows:
	\begin{eqnarray}
		C_{22}s_d^2+C_{2sc}s_dc_d+C_{20}=0
		\label{Eq.sumruleeq02}
	\end{eqnarray}
with the coefficients as follows:
	\begin{eqnarray*}
		C_{22}&=&-2(\cos(\lambda^u_2)-1)(8c_u^2-12s_u^2-4\sqrt{3}c_us_u)
			+36(s_u^2-c_u^2)
		\\
		C_{2sc}&=&-2(\cos(\lambda^u_2)-1)(-12c_us_u+4\sqrt{3}c_u^2)
			+72c_us_u
		\\
		C_{20}&=&-2(\cos(\lambda^u_2)-1)(3-8c_u^2+4\sqrt{3}c_us_u)
			+36c_u^2
			\\
			&&-36\Big(c_{12}^2c_{23}^2+s_{13}^2s_{23}^2s_{12}^2-2c_{12}s_{12}c_{23}s_{23}s_{13}\cos[\delta_{CP}]\Big)
	\end{eqnarray*}
Eliminating $\theta^d$ from Eqs. (\ref{Eq.sumruleeq01}) and (\ref{Eq.sumruleeq02}), we obtain
	\begin{eqnarray}
		&&C_{12}^2C_{20}^2
		+C_{1sc}^2C_{20}^2
		-2C_{10}C_{20}C_{12}C_{22}
		+C_{1sc}^2C_{20}C_{22}
		+C_{10}^2C_{22}^2
		-2C_{10}C_{20}C_{1sc}C_{2sc}
		\nonumber\\
		&&-C_{12}C_{20}C_{1sc}C_{2sc}
		-C_{10}C_{22}C_{1sc}C_{2sc}
		+C_{10}^2C_{2sc}^2
		+C_{10}C_{12}C_{2sc}^2=0
	\label{Eq.sumruleMid01}
	\end{eqnarray}
For the given CKM mixing angles and CPV, the above formula is only dependent on the parameters $\theta^u,\lambda^u_2$.

Taking $\lambda_1^u=0$ in Eqs. (\ref{Eq.s130}) and (\ref{Eq.s230}), we have
\begin{eqnarray*}
	s_{13}^2
		=\frac{2}{9}\sin^2(\frac{\lambda^u_2}{2})(\sqrt{3}c_u+s_u)^2
		\\
		\frac{s_{23}^2}{1-s_{13}^2}
		=\frac{2}{9}\sin^2(\frac{\lambda^u_2}{2})(c_u-\sqrt{3}s_u)^2
	\end{eqnarray*}
$\theta^u$ and $\lambda^u_2$ can be solved as
	\begin{eqnarray*}
		\pm\sqrt{1-\cos(\lambda^u_2)}c_u&=&\frac{3}{4}\Big(\sqrt{3}s_{13}+\frac{s_{23}}{\sqrt{1-s_{13}^2}}\Big)
		\\
		\pm\sqrt{1-\cos(\lambda^u_2)}s_u&=&\frac{3}{4}\Big(s_{13}-\frac{\sqrt{3}s_{23}}{\sqrt{1-s_{13}^2}}\Big).
	\end{eqnarray*}
Eliminating $\lambda^u_2$ and $\theta^u$ in Eq. (\ref{Eq.sumruleMid01}), a complicated flavor mixing sum rule is obtained.

\begin{figure}[htbp]
\begin{center}
\includegraphics[height=0.3 \textheight]{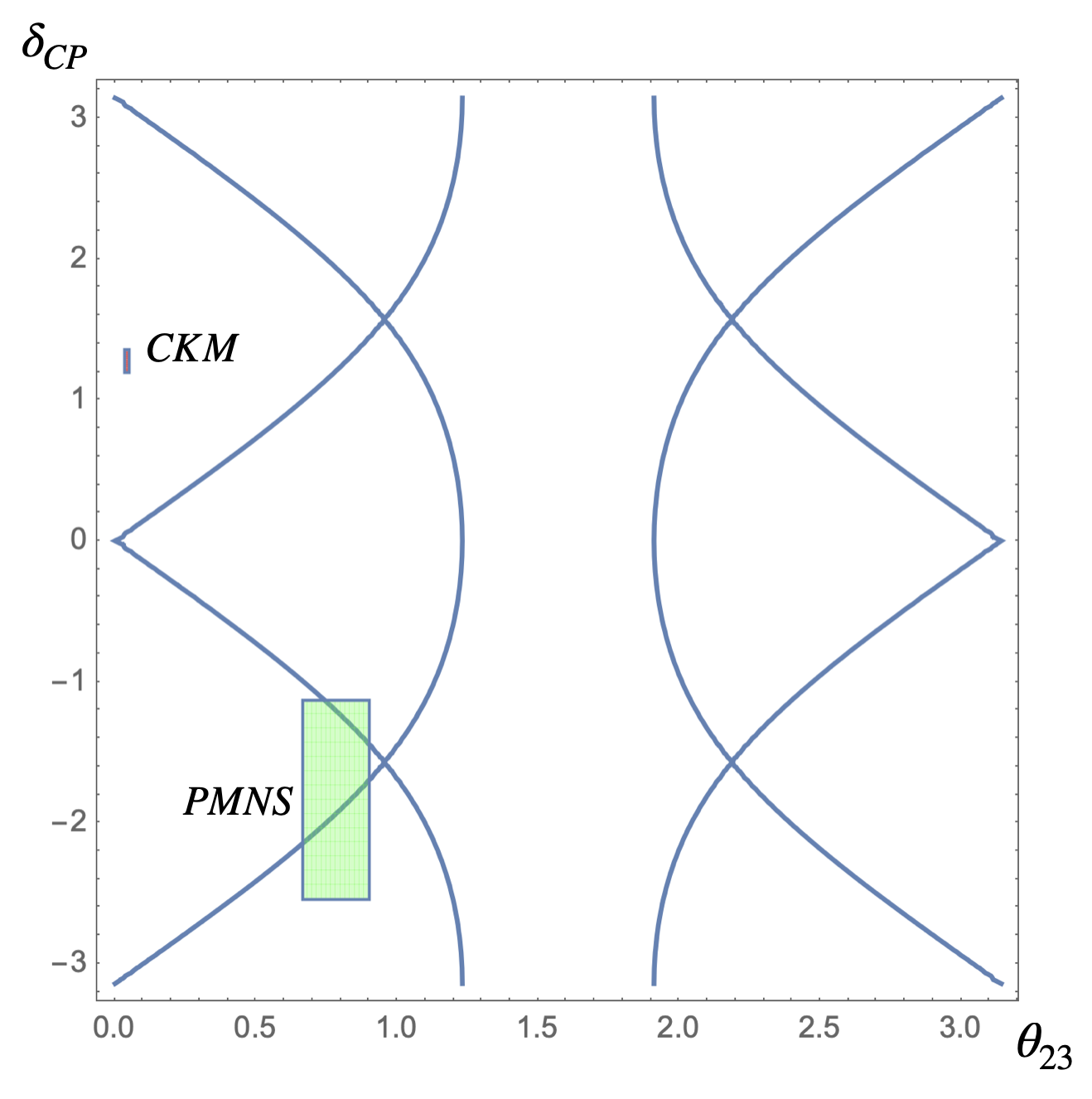}
\caption{A prediction from the flavor mixing sum rule for nonvanishing $s_{13}$ in hierarchy limit. The range of $\theta_{23}-\delta_{CP}$ for the CKM (red) and PMNS (green) is in 1$\sigma$ C.L.}
\label{fig.sumrule}
\end{center}
\end{figure}

To investigate the sum rule more clearly, we expand it in terms of the power of $s_{13}$. The leading order appears as the order of $s_{13}^2$
	\begin{eqnarray*}
		&&s_{12}^2c_{12}^2s_{23}^2 s_{13}^2
		\Big(
		1
		-6c_{23}^2
		-4 \cos^2\delta_{CP}c_{23}^2
		+9c_{23}^4\Big)
		+\mathcal{O}(s_{13}^3)=0.
	\end{eqnarray*}
The result shows that there are two possible cases:
	\begin{itemize}
		\item[{\bf case 1:}] $s_{13}=0$. This makes the sum rule valid up to any power of $s_{13}$. This case arises from the condition of the mass hierarchy limit, which is consistent with the interpretation of the smallness of $s_{13}$ in the CKM (and in PMNS).
		\item[{\bf case 2:}] $s_{13}\neq0$. For nonvanishing $s_{12}$ and $s_{23}$, the sum rule predicts a simple relation between $\theta_{23}$ and $\delta_{CP}$ as follows:
$$1-6c_{23}^2-4 c_{23}^2\cos^2(\delta_{CP})+9c_{23}^4=0.$$
Regarding CKM, this relation is not inadaptable. However, it cannot be ruled out for the PMNS due to the large experimental deviation of $\delta_{CP}$ (see Fig. \ref{fig.sumrule} for details).
\end{itemize}


\section{Unified Yukawa Interactions}\label{sec.unifiedYukawa}
In the MFS, a flat mass matrix is a corollary of Hypothesis II. In fact, there exists a corresponding relation between the hierarchal masses and flat structures. When enlarging the range of the research objects, more hierarchies can be widely found between up-type and down-type total quark masses $m^u_{\Sigma}\gg m^d_{\Sigma}$, charged lepton and neutrino total mass $m^e_{\Sigma}\gg m^\nu_{\Sigma}$, and even total quark mass and total lepton mass $m^u_{\Sigma}+m^d_{\Sigma}\gg m^\nu_{\Sigma}+m^e_{\Sigma}$. Thus, a unified Yukawa interaction with a single coupling can be established along the flat structure.

\subsection{Uptype and Downtype Quarks}\label{sec.quarkleptonYuk}
In family universal Yukawa terms in Eq. (\ref{Eq.FamilyUniversalYukawaTerm}), a flat matrix is introduced to represent a nondistinguishing Yukawa interaction between the different families in the hierarchy limit.

To build a unified Yukawa interaction for up-type and down-type quarks, we consider two quark fields
$\psi^{q1},\psi^{q2}$ in the Yukawa basis (for the sake of convenience, the superscript $^{(Y)}$ of the field has been neglected).
Their left-handed components form a $SU(2)_L$ doublet $\Psi_L^q=(\psi^{q1}_L,~ \psi^{q2}_L)^T$, and the right-handed $\psi^{q1}_R$ and $\psi^{q2}_R$ remain singlets. Considering $SU(2)_L$ doublet scalars $\Phi$, $\bar{\Psi}_L^q\Phi$ and $\bar{\Psi}_L^q\tilde{\Phi}$ form $SU(2)_L$ singlets.
We assume that there exists a nondistinguishing Yukawa interaction between $SU(2)_L$ invariants: $\bar{\Psi}_L^q\Phi\psi^{q1}_R$, $\bar{\Psi}_L^q\tilde{\Phi}\psi^{q1}_R$, $\bar{\Psi}_L^q\Phi\psi^{q2}_R$ and $\bar{\Psi}_L^q\tilde{\Phi}\psi^{q2}_R$; that is, there is a flat structure as follows:
	\begin{eqnarray}
				-\mathcal{L}^Q_Y=\frac{1}{2}y^q\bar{\Psi}_L^q\left(\tilde{\Phi}, \Phi\right)\Array{cc}{1&1 \\ 1& 1}\Array{c}{\psi^{q1}_R\\ \psi^{q2}_R}
				\label{Eq.quarkYukawaMtx}
	\end{eqnarray}
Here, $y^q$ is a real coupling that determines the quark total mass $m_\Sigma^u+m_\Sigma^d$.
Alternatively, Eq. (\ref{Eq.quarkYukawaMtx}) can also be written into a unified quark Yukawa interaction
\begin{eqnarray}
				-\mathcal{L}^Q_Y=\frac{1}{2}y^q\bar{\Psi}_L^q(\Phi+\tilde{\Phi})(\psi^{q1}_R+\psi^{q2}_R)
				\label{Eq.quarkYukawa1}
			\end{eqnarray}
Here, the scalar $\Phi+\tilde{\Phi}$ is the $SU(2)_L$ self-charged field, i.e., $i\sigma_2(\Phi+\tilde{\Phi})=\Phi+\tilde{\Phi}$.
Notably, in the above unified quark Yukawa interaction, $\psi^{q1},\psi^{q^2}$ are not up-type or down-type quarks but their linear combinations.
We make an orthogonal rotation as follows:
	\begin{eqnarray}
		\Psi_L^q&\rightarrow& {R}_{ud}\Psi_L^q=\Array{c}{u^{(Y)}_{L} \\ d^{(Y)}_{L}}
		\\
		\Array{c}{\psi^{q1}_{R} \\ \psi^{q2}_{R}}&\rightarrow &{R}_{ud}\Array{c}{\psi^{q1}_{R} \\ \psi^{q2}_{R}}=\Array{c}{u^{(Y)}_{R} \\ d^{(Y)}_{R}}
		\\
	\left(\tilde{\Phi},\Phi\right)&\rightarrow& {R}_{ud}\left(\tilde{\Phi},\Phi\right)_{2\times 2}{R}_{ud}^T=\left(\tilde{H},H\right)
	\label{Eq.Rud}
	\end{eqnarray}
with
\begin{eqnarray}
		{R}_{ud}\equiv\frac{1}{\sqrt{2}}\Array{cc}{1& 1 \\ -1 & 1},
		\label{Eq.Rud}
	\end{eqnarray}
Moreover, the flat matrix in Eq. (\ref{Eq.quarkYukawaMtx}) is diagonalized into the family-universal form as shown in Eq. (\ref{Eq.FamilyUniversalYukawaTerm})
	\begin{eqnarray}
				-\mathcal{L}_Y^Q
				=\bar{Q}^{(Y)}_L \Array{cc}{\tilde{H}&  H}\Array{cc}{y^u & \\ & y^d}\Array{c}{u^{(Y)}_R \\ d^{(Y)}_R}
				\label{Eq.UnifiedquarkLag}
			\end{eqnarray}
In the hierarchy limit $m_\Sigma^u\gg m_\sigma^d$, Yukawa couplings have the values $y^d=0$ and $y^u=y^q$. This means that the total quark mass $m_\Sigma^u+m_\Sigma^d$ is occupied by up-type quarks.

Regarding a more realistic case, we consider the flat structure that breaks by a perturbation in a nondiagonal element such as
\begin{eqnarray}
		\Array{cc}{1&1 \\ 1& 1}\rightarrow\Array{cc}{1&1+\delta^q \\ 1+\delta^q& 1}\equiv {I}_\delta^q.
		\label{Eq.deltaq}
	\end{eqnarray}
By the same rotation as Eq. (\ref{Eq.Rud}), ${I}_\delta^q$ is diagonalized into hierarchal eigenvalues with $y^u\gg y^d$, i.e.,
$$\frac{y^q}{2}{R}_{ud}{I}_\delta^{q}{R}^T_{ud}=\text{diag}(y^u,y^d).$$
To yield a realistic $m_\Sigma^u$ and $m_\Sigma^d$, $\delta_q$ must satisfy
\begin{eqnarray*}
	\frac{y^d}{y^u}=\frac{-\delta^q}{2+\delta^{q}}.
			\end{eqnarray*}
Its value is fixed as
$$\delta^{q}=-\frac{2m^d_{\Sigma}}{m^u_{\Sigma}}\simeq-\frac{2m^b}{m^t}\simeq{-0.0483}.$$

Similarly, unified Yukawa interactions can be generalized to leptons. We consider two lepton fields $\psi^{l1}$ and $\psi^{l2}$. Their left-handed components form a $SU(2)_L$ doublet $\Psi_L^l=(\psi^{l1}_L,~ \psi^{l2}_L)^T$ and right-handed $\psi^{l1}_R$, and $\psi^{l2}_R$ maintain singlets.
A flat structure is assumed for nondistinguishing Yukawa interactions: $\bar{\Psi}_L^l\Phi\psi^{l1}_R$, $\bar{\Psi}_L^l\tilde{\Phi}\psi^{l1}_R$, $\bar{\Psi}_L^l\Phi\psi^{l2}_R$ and $\bar{\Psi}_L^l\tilde{\Phi}\psi^{l2}_R$.
Neglecting the similar scenario as quarks, a difference between the quarks and leptons originates from the total mass hierarchy order. Because $m_\Sigma^\nu\ll m_\Sigma^e$ differs from $m_\Sigma^u\gg m_\Sigma^d$, the diagonalizing rotation for leptons is required as follows:
$${R_{\nu e}}=\frac{1}{\sqrt{2}}\Array{cc}{1& -1 \\ 1 & 1}.$$
To yield the realized lepton mass (for $m_1^\nu=0.001$ eV), a nondiagonal perturbation for leptons must satisfy the following condition:
$\delta^{l}=-\frac{2y^\nu}{y^\nu+y^e}$.

Moreover, a unified quark/lepton Yukawa interaction is expressed as follows:
\begin{eqnarray}
			-\mathcal{L}^{Q+L}_Y=\frac{1}{2}y^q\bar{\Psi}^{q}_L(\Phi+\tilde{\Phi})(\psi^{q1}_R+\psi^{q2}_R)
				+\frac{1}{2}y^l\bar{\Psi}^{l}_L(\Phi+\tilde{\Phi})(\psi^{l1}_R+\psi^{l2}_R).
			\label{Eq.YukLagQuarkLepton}
		\end{eqnarray}
with two universal couplings: $y^q$ and $y^l$.

\subsection{Quarks and Lepton}\label{sec.unifiedYuk}
Inspired by the mass hierarchy between leptons and quarks $m_\Sigma^u+m_\Sigma^d\gg m_\Sigma^\nu+m_\Sigma^e$, we built a flat structure in a similar way.

Considering four fermions $\psi^{Ai},\psi^{Bi}$ for $i=1,2$, their left-handed components are combined into two $SU(2)_L$ doublets:
	\begin{eqnarray*}
	\Psi_{L}^A=\Array{c}{\psi^{A1}_L \\ \psi^{A2}_L},~~\Psi_{L}^B=\Array{c}{\psi^{B1}_L\\ \psi^{B2}_L}
	\end{eqnarray*}
where $\psi^{A1}_R,\psi^{A2}_R,\psi^{B1}_R$ are four right-handed values, and $\psi^{B2}_R$ remain singlets.
If nondistinguishing Yukawa interactions exist, we obtain:
	\begin{eqnarray*}
		&&\Psi_L^A(\Phi+\tilde{\Phi})(\psi_R^{A1}+\psi_R^{A2}),~~
		\Psi_L^A(\Phi+\tilde{\Phi})(\psi_R^{B1}+\psi_R^{B2}),~~
		\\
		&&\Psi_L^B(\Phi+\tilde{\Phi})(\psi_R^{A1}+\psi_R^{A2}),~~
		\Psi_L^B(\Phi+\tilde{\Phi})(\psi_R^{B1}+\psi_R^{B2}),~~
	\end{eqnarray*}
These terms can be arranged into a flat matrix as follows:
	\begin{eqnarray}
				-\mathcal{L}_Y
				=\frac{y^0}{4}\left(\bar{\Psi}^{A}_L, \bar{\Psi}^{B}_L\right)(\Phi+\tilde{\Phi})\Array{cc}{1&1\\ 1&1}\Array{c}{\sum_i\psi^{Ai}_R \\ \sum_i\psi^{Bi}_R}
				\label{Eq.UniversalYukawaMtx}
			\end{eqnarray}
Compactly, $\mathcal{L}_Y$ can be expressed as a unified Yukawa interaction between the left-handed doublet $(\bar{\Psi}^{A}_L+ \bar{\Psi}^{B}_L)$, right-handed singlet $\psi^{Ai}_R +\psi^{Bi}_R$ and self-conjugation scalar $(\Phi+\tilde{\Phi})$
as follows:
	\begin{eqnarray}
				-\mathcal{L}_Y
				=\frac{y^0}{4}\left(\bar{\Psi}^{A}_L+ \bar{\Psi}^{B}_L\right)(\Phi+\tilde{\Phi})\sum_{i=1,2}\left(\psi^{Ai}_R +\psi^{Bi}_R\right)
				\label{Eq.UniversalYukawa1}
			\end{eqnarray}
where $y^0$ is a single coupling, which determines all fermion total masses $m_\Sigma^u+m_\Sigma^d+m_\Sigma^\nu+m_\Sigma^e$.

After diagonalizing the flat matrix in Eq. (\ref{Eq.UniversalYukawaMtx}) by a rotation $\bf {R}_{lq}=\frac{1}{\sqrt{2}}\Array{cc}{1&-1\\ 1 & 1}$, leptons and quarks can be divided as follows:
	\begin{eqnarray}
		\Array{c}{\Psi^{A}_L \\ \Psi^{B}_L}&\rightarrow& {R}_{lq}\Array{c}{\Psi^{A}_L \\ \Psi^{B}_L}=\Array{c}{\Psi^{l}_L \\ \Psi^{q}_L}
		\label{Eq.PsiAB01}
		\\
		\Array{c}{\sum_i\psi^{Ai}_R \\ \sum_i\psi^{Bi}_R}&\rightarrow& {R}_{lq}\Array{c}{\sum_i\psi^{Ai}_R \\ \sum_i\psi^{Bi}_R}=\Array{c}{\sum_i\psi^{li}_R \\ \sum_i\psi^{qi}_R}.
		\label{Eq.PsiAB02}
	\end{eqnarray}
Quark Yukawa coupling $y^q$ and lepton coupling $y^l$ are obtained as follows:
	\begin{eqnarray}
	\frac{y^0}{2}\Array{cc}{1& 1\\ 1& 1}\rightarrow \frac{y^0}{2}{R}_{lq}{I}_\delta^s{R}^T_{lq}=\Array{cc}{y^l & \\ & y^q}.
	\label{Eq.YukQLdiag}
	\end{eqnarray}
and Eq. (\ref{Eq.UniversalYukawaMtx}) becomes the unified quark/lepton Yukawa terms, as shown in Eq. (\ref{Eq.YukLagQuarkLepton}).

To address the realized case, a nondiagonal perturbation $\delta^s$ is needed.
	\begin{eqnarray}
	\Array{cc}{1 & 1\\ 1 & 1}\rightarrow \Array{cc}{1 & 1+\delta^s \\ 1+\delta^s &1}\equiv {I}_\delta^s.
	\label{Eq.YukStrongCorr}
	\end{eqnarray}
By the same rotation transformation $\bf {R}_{lq}$, Eq. (\ref{Eq.YukQLdiag}) yields the physical $y^q$ and $y^l$ under the condition of
	\begin{eqnarray*}
	\delta_{S}=-\frac{2(m^\nu_{\Sigma}+m^e_{\Sigma})}{m^u_{\Sigma}+m^d_{\Sigma}}\simeq-\frac{2m^\tau}{m^t}\simeq {-0.0205}.
	\end{eqnarray*}

To date, a unified Yukawa interaction for SM fermions has been built in Eq. (\ref{Eq.UniversalYukawa1}) in the hierarchy limit.
When a perturbation is introduced into the flat matrix, the unified Yukawa term is broken into the realized case at different levels.
Regarding a possible origin of perturbations $\delta^q$, $\delta^l$ and $\delta^s$, some clues may be found by comparing the cases before and after breaking.
In Eqs. (\ref{Eq.quarkYukawaMtx}) and (\ref{Eq.UnifiedquarkLag}), linear combined quarks $\psi^{q1}$ and $\psi^{q2}$ are broken into up-type quark $u$ and down-type quark $d$. Since $u$ has an electric charge of $+2/3$ and $d$ has an electric charge of $-1/3$, electric charge symmetry is maintained. Therefore, $\delta^q$ may arise from a quantum correction of quarks with different electric charges. This case is also suitable for linear combined leptons $\psi^{l1}$ and $\psi^{l2}$ broken to neutrinos with an electric charge of $0$ and charged leptons with an electric charge of $-1$. At the higher level, $\Psi^A$ and $\Psi^B$ are broken into quarks and leptons in Eqs. (\ref{Eq.PsiAB01}) and (\ref{Eq.PsiAB02}), in which the color symmetry is maintained due to quarks with color charges and leptons without color charges. This suggests that $\delta^s$ in Eq. (\ref{Eq.YukStrongCorr}) may arise from strong interactions.

\begin{figure}
	\centering
	\includegraphics[height=0.27\textheight]{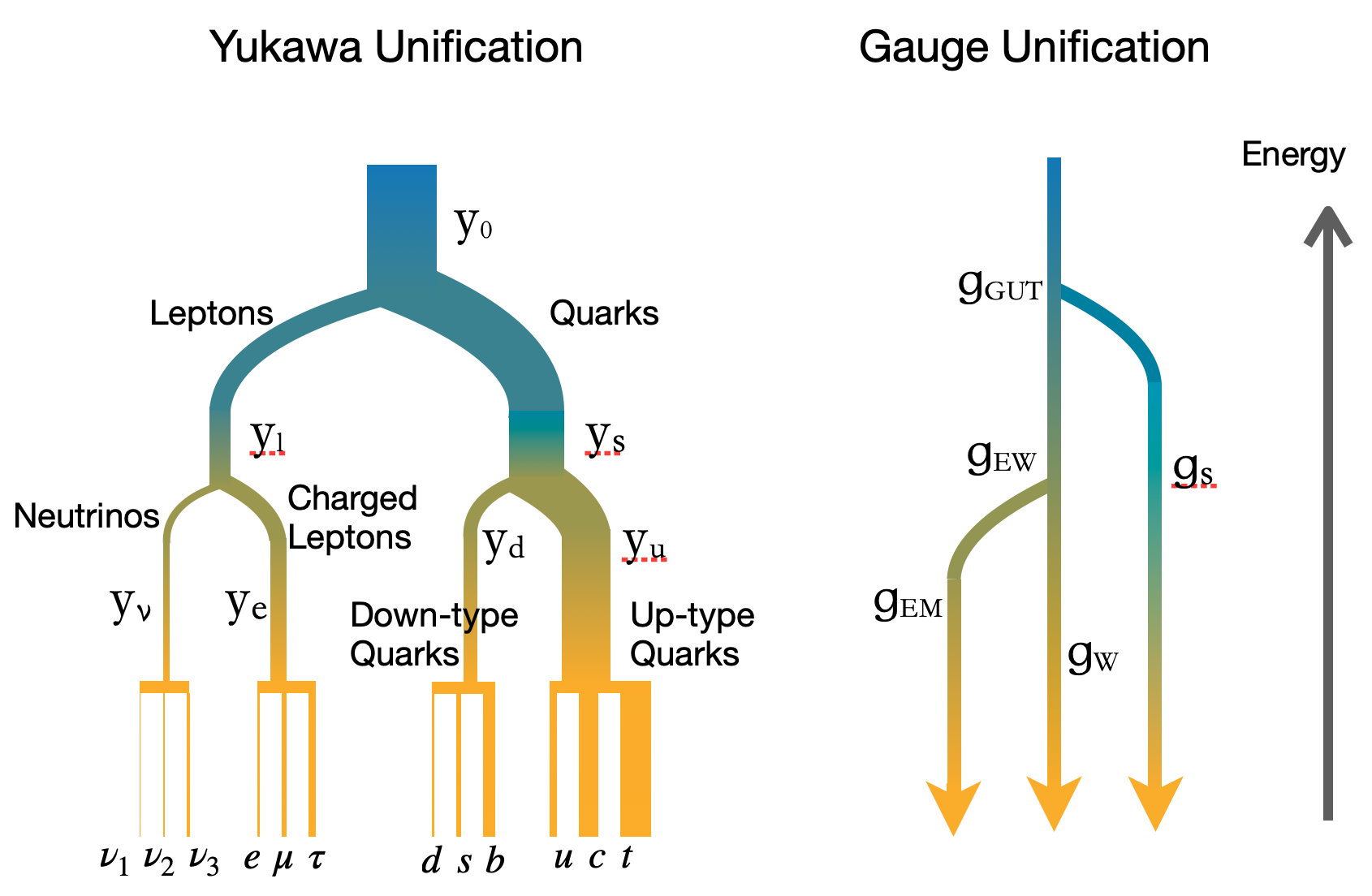}
	\caption{Gauge interaction unification and Yukawa interaction unification.}
	\label{fig.unifyYuk}
\end{figure}

The unified Yukawa interaction provides a compacted and desired scenario to comprehend the fermion flavor structure. This scenario is similar to the gauge unification in the GUT, as shown in Fig. \ref{fig.unifyYuk}. At the GUT scale, strong and EW interactions are separated. Fermions taking part in strong interactions become quarks, while others become leptons. Regarding the EW scale, differential EM interaction separates quarks into up-type and down-type interactions and separates leptons into neutrinos and charged leptons.

\section{Summary}\label{sec.summary}
Through the redundancy of the SM Yukawa couplings and similarity of quark and lepton mixings, a family universal Yukawa interaction has been proposed for both quarks and leptons with exactly the same number of parameters as the number of phenomenological observables.
The MFS answers the question of unclear flavor structures in the SM according to a mismatch of gauge eigenstates in Yukawa terms.
Regarding the Yukawa basis, the CP violation in the CKM/PMNS can stem from the phase transformation between the Yukawa basis and gauge basis.
Yukawa interactions between different families have a close-to-flat flavor structure.
With the help of the quark and lepton mass hierarchy, a $[SO(2)_L\times SO(2)_R]^f$ symmetry can be found in the flat mass matrix. When hierarchy corrections are considered, the symmetry is broken into $SO(2)_{LR}^f$, which yields a prominent result in the quark/lepton mixing matrix. Research shows that the CKM/PMNS is dominated by only 4 parameters: 2 Yukawa phases and 2 $SO(2)$ rotation angles, which are independent of the mass ratios of quarks/leptons. The validity of the MFS is verified by fitting the quark/lepton masses and mixing parameters.

Based on the role of the Yukawa phase $\lambda_1^{u,\nu}$ in the mixing matrix, a sum rule on the mixing parameters is deduced. The sum rule required nonvanishing $s_{13}$ or a simple constraint relation about $\theta_{23}$ and $\delta_{CP}$ in the hierarchy limit. This suggests that the smallness of $s_{13}$ in the CKM and PMNS may originate from the mass hierarchy.

The flat mass matrix can be generalized to two kinds of quarks, to two kinds of leptons and to all fermions since the mass hierarchy widely exists at multiple levels. Although actual fermion masses can be achieved through nondiagonal corrections in a flat mass matrix, the dynamics explanation of these corrections, $\delta^q,\delta^l$ and $\delta^s$, is still unknown. Discovering their origins will provide a deeper understanding of the fermion flavor structure in the future.
\section*{Acknowledgments}
This work is supported by Shaanxi Natural Science Foundation of China (2022JM-052).

\begin{appendix}
\section{Hierarchy Corrections to CKM Mixing Parameters}
In the hierarchy limit, the CKM matrix has a structure as shown in Eq. (\ref{Eq.UckmS00}), and the mixing angles and Jarlskog invariant can be calculated from Eq. (\ref{Eq.s130}), (\ref{Eq.s230}), (\ref{Eq.s120}) and (\ref{Eq.Jcp0}). When flavor breaking is considered, the CKM matrix receives corrections from $h_{23}^{u,d}$, while its structure is maintained in Eq. (\ref{Eq.ClostToFlatCKM01}) or (\ref{Eq.ClostToFlatCKM02}). Hierarchy corrections can be defined as
$$\Delta s_{ij}^2=s_{ij}^2\Big|_h-s_{ij}^2\Big|_0$$
Here, the subscript $_h$ labels a mixing parameter with hierarchy corrections, and the subscript $_0$ labels that in the hierarchy limit. In $\mathcal{O}(h_{ij}^1)$, these results can be expressed as
\begin{eqnarray}
		\Delta s_{13}^2&=&C_{13}^uh_{23}^u+C_{13}^dh_{23}^d+\mathcal{O}(h^2)
		\label{Eq.Deltas13h1}\\
		\Delta s_{12}^2&=&C_{12}^uh_{23}^u+C_{12}^dh_{23}^d+\mathcal{O}(h^2)
		\label{Eq.Deltas12h1}\\
		\Delta s_{23}^2&=&C_{23}^uh_{23}^u+C_{23}^dh_{23}^d+\mathcal{O}(h^2)
		\label{Eq.Deltas23h1}
	\end{eqnarray}
The above coefficients $C_{ij}^{u,d}$ are listed as follows:
\begin{eqnarray*}
	C_{13}^u&=&-\frac{1}{6} s_u(c_1+c_2+1) \Big[-2c_1 s_u+\sqrt{3} (c_2-1) c_u+ c_2 s_u+ s_u\Big]
		\\
		&&-\frac{1}{6} s_u(s_1+ s_2)(-2 s_1 s_u+ s_2 s_u+\sqrt{3} s_2 c_u)
\end{eqnarray*}
\begin{eqnarray*}
	C_{13}^d&=&-\frac{1}{36}\Big[s_u (-2 c_1+c_2+1)-\sqrt{3}(1-c_2)\Big]
		\\
		&&~~~\times\Big\{s_u \Big[4 c_1(2 c_d-3)+c_2 (-2 \sqrt{3} s_d+2 c_d-3)+2 \sqrt{3} s_d+2 c_d-3\Big]
		\\
		&&~~~~~~+c_u \Big[c_2 (2 \sqrt{3} c_d-6 s_d-3\sqrt{3})-6 s_d-2 \sqrt{3} c_d+3 \sqrt{3}\Big]\Big\}
	\\
	&&-\frac{1}{36 }\Big[s_u (s_2-2 s_1)+\sqrt{3} s_2 c_u\Big] 
	\\
	&&~~~\times\Big\{s_u \Big[4 s_1 (2 c_d-3)+s_2 (-2 \sqrt{3} s_d+2 c_d-3)\Big]
	+s_2 c_u \Big[2 \sqrt{3} c_d-3 (2 s_d+\sqrt{3})\Big]\Big\}
\end{eqnarray*}
\begin{eqnarray*}
	C_{12}^u&=&\frac{1}{216}s_u \Big[-[s_u (-2 c_1+c_2+1)-\sqrt{3} (1-c_2) c_u]^2
		-[s_u (s_2-2 s_1)+\sqrt{3} s_2 c_u]^2+18\Big] 
		\\
		&&~~~\times \Big\{
			\Big[c_d (2 c_1-c_2-1)+\sqrt{3} (c_2-1) s_d\Big] 
			\Big[c_d [s_u (4 c_1+c_2+1)- \sqrt{3} (1-c_2) c_u]
			\\
			&&~~~~~~-s_d [3 (c_2+1) c_u+\sqrt{3} (c_2-1) s_u]\Big]
		\\
		&&~~~~~~+\Big[c_d (2 s_1-s_2)+\sqrt{3} s_2 s_d\Big] \Big[c_d [s_u (4 s_1+s_2)+\sqrt{3} s_2 c_u]-s_2 s_d (\sqrt{3} s_u+3 c_u)\Big]\Big\}
		\\
		&&-\frac{1}{216}s_u \Big\{c_1 \Big[\sqrt{3} (1-c_2) c_u+(c_2+1) s_u\Big]+s_u (s_1 s_2-2 c_2)-\sqrt{3} s_1 s_2 c_u\Big\}
		\\
		&&~~~\times \Big\{\Big[c_d (\sqrt{3} (1-c_2) c_u-s_u (4 c_1+c_2+1))+s_d (3 (c_2+1) c_u+\sqrt{3} (c_2-1) s_u)\Big]^2
		\\
		&&~~~~~~+\Big[s_2 s_d (\sqrt{3} s_u+3 c_u)-c_d (s_u (4 s_1+s_2)+\sqrt{3} s_2 c_u)\Big]^2\Big\}
\end{eqnarray*}
\begin{eqnarray*}
	C_{12}^d&=&\frac{1}{648} (3 c_d-2) \Big\{\Big[s_u (-2 c_1+c_2+1)-\sqrt{3} (1-c_2) c_u\Big]^2
			\\
			&&~~~~~~~~~+\Big[s_u (s_2-2 s_1)+\sqrt{3} s_2 c_u\Big]^2-18\Big\}
		\\
		&&~~~\times \bigg\{\Big[s_u (-2 c_1+c_2+1)-\sqrt{3} (1-c_2) c_u\Big] 
			\Big[c_d[s_u (4 c_1+c_2+1)-\sqrt{3} (1-c_2) c_u]
			\\
			&&~~~~~~~~~-s_d (3 (c_2+1) c_u+\sqrt{3} (c_2-1) s_u)\Big]
			+\Big[s_u (s_2-2 s_1)+\sqrt{3} s_2 c_u\Big] 
			\\
			&&~~~~~~~~~\times\Big\{s_uc_d  (4 s_1+s_2)+\sqrt{3} s_2 c_uc_d 
			-s_2 s_d (\sqrt{3} s_u+3 c_u)\Big\}\bigg\}
		\\
		&&+\frac{\sqrt{2}}{2592} \Bigg\{
			\Big[\sqrt{2} s_u (-2 c_1+c_2+1)-\sqrt{6} (1-c_2) c_u\Big]
			 \\
			 &&~~~\times\Big\{s_u \Big[4 c_1 (2 c_d-3)+c_2 (-2 \sqrt{3} s_d+2 c_d-3)+2 \sqrt{3} s_d+2 c_d-3\Big]
			 	\\
				&&~~~~~~+c_u \Big[c_2 (2 \sqrt{3} c_d-3 (2 s_d+\sqrt{3}))-6 s_d-2 \sqrt{3} c_d+3 \sqrt{3}\Big]\Big\}
			\\
			&&~~~~~~+\Big[\sqrt{2} s_u (s_2-2 s_1)+\sqrt{6} s_2 c_u\Big]
			\Big\{s_u \Big[4 s_1 (2 c_d-3)+s_2 (-2 \sqrt{3} s_d+2 c_d-3)\Big]
			\\
			&&~~~~~~~~~+s_2 c_u \Big[2 \sqrt{3} c_d-3 (2 s_d+\sqrt{3})\Big]\Big\}
			\Bigg\}
		\\
		&&~~~\times \Bigg\{\Big[c_d [\sqrt{3} (1-c_2) c_u-s_u (4 c_1+c_2+1)]
			+s_d [3 (c_2+1) c_u+\sqrt{3} (c_2-1) s_u]\Big]^2
		\\
		&&~~~~~~+\Big[s_2 s_d (\sqrt{3} s_u+3 c_u)-c_d (s_u (4 s_1+s_2)+\sqrt{3} s_2 c_u)\Big]^2\Bigg\}
\end{eqnarray*}
\begin{eqnarray*}
	C_{23}^u&=&\frac{1}{162}(3 c_u-2) 
			\Big[c_1 [(c_2+1) c_u+\sqrt{3} (c_2-1) s_u]
			+s_1 s_2 (\sqrt{3} s_u+c_u)-2 c_2 c_u\Big]
		\\
		&&~~~\times\Big[-2\sqrt{3} c_1 (1-c_2) s_u c_u+c_2 (2 c_1 \sin^2(\theta^u)+2 \cos(2 \theta^u)+1)
			\\
			&&~~~~~~+2 s_1 s_2 s_u (s_u+\sqrt{3} c_u)+2 c_1 \sin^2(\theta^u)+6\Big]
		\\
		&&-\frac{1}{54}s_u \Big[s_1 s_2(-1+2c_u^2-2 \sqrt{3} s_u c_u)
			+2 c_1 c_u [(c_2+1) c_u-\sqrt{3} (c_2-1) s_u]
			\\
			&&~~~~~~+s_1 s_2
			+4 c_2 s_u^2-c_2-3\Big] 
			\Big\{-c_1 [\sqrt{3} (1-c_2) c_u+(c_2+1) s_u]
			\\
			&&~~~~~~+s_u (2 c_2-s_1 s_2)+\sqrt{3} s_1 s_2 c_u\Big\}
\end{eqnarray*}
\begin{eqnarray*}
	C_{23}^d&=&\frac{1}{648}\Big[18- [ s_u (-2 c_1+c_2+1)-\sqrt{3} (1-c_2) c_u]^2
		- [ s_u (s_2-2 s_1)+\sqrt{3} s_2 c_u]^2\Big] 
		\\
		&&~~~\times \bigg\{ \Big[c_u (2 c_1-c_2-1)+\sqrt{3} (c_2-1) s_u\Big] 
			\Big\{c_u (4 c_1 (2 c_d-3)
			\\
			&&~~~~~~+c_2 (-2 \sqrt{3} s_d+2 c_d-3)+2 \sqrt{3} s_d+2 c_d-3)
			\\
			&&~~~~~~+s_u \Big[c_2 (6 s_d-2 \sqrt{3} c_d+3 \sqrt{3})+6 s_d+2 \sqrt{3} c_d-3 \sqrt{3}\Big]\Big\}
			\\
			&&~~~~~~+[c_u (2 s_1-s_2)+\sqrt{3} s_2 s_u]
			\Big[4 s_1 (2 c_d-3) c_u
				\\
				&&~~~~~~-s_2 (2 \sqrt{3} s_uc_d+2 \sqrt{3} c_us_d
				-2 c_uc_d-6s_us_d
				-3 \sqrt{3} s_u+3 c_u)\Big]\bigg\}
		\\
		&&+\frac{1}{648}\Big\{\Big[c_u (2 c_1-c_2-1)+\sqrt{3} (c_2-1) s_u\Big]^2
			+\Big[c_u (2 s_1-s_2)+\sqrt{3} s_2 s_u\Big]^2\Big\}
		\\
		&&~~~\times \Bigg[\Big[s_u (-2 c_1+c_2+1)-\sqrt{3} (1-c_2) c_u\Big]
			\Big\{s_u \Big[4 c_1 (2 c_d-3)
			\\
			&&~~~~~~+c_2 (-2 \sqrt{3} s_d+2 c_d-3)+2 \sqrt{3} s_d+2 c_d-3\Big]
			\\
			&&~~~~~~+c_u\Big[c_2 (2 \sqrt{3} c_d-3 (2 s_d+\sqrt{3}))-6 s_d-2 \sqrt{3} c_d+3 \sqrt{3}\Big]\Big\}
		\\
		&&~~~~~~+\Big[s_u (s_2-2 s_1)+\sqrt{3} s_2 c_u\Big] 
			\Big\{s_u \Big[4 s_1 (2 c_d-3)+s_2 (-2 \sqrt{3} s_d+2 c_d-3)\Big]
			\\
			&&~~~~~~+s_2 c_u \Big[2 \sqrt{3} c_d-3 (2 s_d+\sqrt{3})\Big]\Big\}\Bigg]
\end{eqnarray*}

Similarly, corrections to the Jarlskog invariant can be written as
\begin{eqnarray*}
		\Delta J_{CP}=C_{J}^uh_{23}^u+C_{J}^dh_{23}^d+\mathcal{O}(h^2).
		\label{Eq.DeltaJcph1}
	\end{eqnarray*}
Coefficients $C_{J}^{u,d}$ are listed as follows:
\begin{eqnarray*}
	C_{J}^u&=& \frac{1}{432}c_1 \sin(2 \lambda_2)\Big\{\sin(2 \theta^d) \Big[2 s_u-9 \sin(2 \theta^u)+4 \sin(3 \theta^u)-3 \sqrt{3} \cos(2 \theta^u)
		\\
		&&~~~+2 \sqrt{3} c_u\Big]
		+\cos(2 \theta^d) \Big[ -2\sqrt{3} s_u+9\sqrt{3} \sin(2 \theta^u)-4\sqrt{3} \sin(3 \theta^u)
		\\
		&&~~~-6 c_u+9 \cos(2 \theta^u)\Big]\Big\}
		\\
		&&+\frac{1}{432}s_1 \cos(2 \lambda_2) \Big\{\sin(2 \theta^d) 
			\Big[-2 s_u+9 \sin(2 \theta^u)-4 \sin(3 \theta^u)-2 \sqrt{3} c_u
			\\
			&&~~~+3 \sqrt{3} \cos(2 \theta^u)\Big]
		+\cos(2 \theta^d) \Big[2\sqrt{3} s_u-9\sqrt{3} \sin(2 \theta^u)+4\sqrt{3} \sin(3 \theta^u)
		\\
		&&~~~+6 c_u-9 \cos(2 \theta^u)\Big]\Big\}
		\\
		&&+\frac{1}{216}s_2\bigg\{2 \sin(2 \theta^d) \Big\{c_1 s_u \Big[\cos(2 \theta^u)+(9-6 \sqrt{3} s_u) c_u-3\Big]
			\\
			&&~~~+2 \cos(2 \lambda_1) [\sin(3 \theta^u)-s_u]
			+\sqrt{3} \Big[5 c_u+3\cos(3 \theta^u)-12 \cos(2 \theta^u)\Big]\Big\}
		\\
		&&~~~- \cos(2 \theta^d) \Big\{3 c_1 \Big[\sqrt{3} [3 s_u-6 \sin(2 \theta^u)+\sin(3 \theta^u)]
		+c_u+3 \cos(2 \theta^u)
		\\
		&&~~~-3 \cos(3 \theta^u)\Big]
		+2 \sqrt{3} (\sin(3 \theta^u)-s_u)\Big\}
		\bigg\}
		\\
		&&+\frac{1}{216} s_1 c_2 \Big\{8 c_1 \sin(2 \theta^d) [s_u-\sin(3 \theta^u)]
			+3\sqrt{3}  \sin(2 (\theta^d-\theta^u))
				\\
				&&~~~-\sqrt{3}\sin(2 \theta^d-\theta^u)
				\\
				&&~~~-6 \sqrt{3}\sin(2 (\theta^d+\theta^u))
				+6\sqrt{3} \sin(2 \theta^d+\theta^u)
				\\
				&&~~~-2\sqrt{3} \sin(2 \theta^d+3 \theta^u)
				-\sqrt{3}\sin(2 \theta^d-3 \theta^u)
			+7 \cos(2 \theta^d-\theta^u)
			\\
			&&~~~+2 \cos(2 \theta^d+\theta^u)
			-2 \cos(2 \theta^d+3 \theta^u)
			-7 \cos(2 \theta^d-3 \theta^u)
		\Big\}
		\\
		&&+\frac{1}{432}s_1 \Big\{12 \cos(2 \theta^d-\theta^u)
			+9 \cos(2 (\theta^d+\theta^u))
			-6 \cos(2 \theta^d+3 \theta^u)
			\\
			&&~~~-12 \cos(2 \theta^d-3 \theta^u)
			+\sqrt{3} \sin(2 \theta^d) [-8 c_u+3 \cos(2 \theta^u)+6 \cos(3 \theta^u)]
			\\
			&&~~~+\sqrt{3} \cos(2 \theta^d) [-16 s_u+27 \sin(2 \theta^u)-2 \sin(3 \theta^u)]\Big\}
\end{eqnarray*}
\begin{eqnarray*}
	C_{J}^d&=&\frac{1}{72} \cos(3 \theta^d) \Big[ \sqrt{3} s_1 (1-c_2) \sin(2 \theta^u)-3 \cos(2 \theta^u) (s_1-c_1 s_2)\Big]
		\\
		&&+\frac{1}{216} c_d \Big\{3 \cos(2 \theta^u) \Big[s_1 (12 c_2 \sin^2(\theta^d)+\cos(2 \lambda_2)+2)-c_1 (s_2+\sin(2 \lambda_2))\Big]
		\\
		&&~~~+\sqrt{3} \sin(2 \theta^u) \Big[s_1 (5 c_2-\cos(2 \lambda_2)-4)+c_1 \sin(2 \lambda_2)+4 s_2\Big]
		\Big\}
		\\
		&&+\frac{1}{216} \cos(2 \theta^u) \Big\{
			2 \sqrt{3} \sin(\frac{\lambda_2}{2}) \Big[s_d (-\cos(\lambda_1-\frac{3 \lambda_2}{2}))
			\\
			&&~~~+9 (c_2+2) \sin(2 \theta^d) \cos(\lambda_1-\frac{\lambda_2}{2})
			-s_1 \sin(\frac{\lambda_2}{2}) [(4 c_2+3) \sin(3 \theta^d)+8 s_d]
			\\
			&&~~~+2 \cos(\frac{\lambda_2}{2}) [s_d-\sin(3 \theta^d)]\Big]
			\\
			&&~~~-c_1 s_2 [\sqrt{3} ((4 c_2+3) \sin(3 \theta^d)+10 s_d)+9 \cos(2 \theta^d)]
		\Big\}
		\\
		&&-\frac{1}{216} \sin(2 \theta^u) \Big\{4 (\sin(3 \theta^d)-s_d) \sin(2 \lambda_1-\lambda_2)
			\\
			&&~~~+2 \cos(\frac{\lambda_2}{2}) \sin(\lambda_1-\frac{\lambda_2}{2}) \Big[(2 c_2-7) s_d+(4 c_2+1) \sin(3 \theta^d)\Big]
			\\
			&&~~~+3 \sqrt{3} (\sin(\frac{3 \lambda_2}{2})-\sin(\frac{\lambda_2}{2})) \cos(2 \theta^d) \cos(\lambda_1-\frac{\lambda_2}{2})
		\Big\}
		\\
		&&+\frac{1}{72} s_2 \Big\{3 \cos(\lambda_1-\lambda_2) \cos(2 (\theta^d+\theta^u))
		\\
		&&~~~+\sin(2 \theta^u) \Big[c_1 (3-2 \sqrt{3} s_d) \sin(2 \theta^d)+4 \sqrt{3} (c_d-2) \cos(2 \theta^d)\Big]\Big\}
\end{eqnarray*}

\end{appendix}

\end{document}